\documentclass[10pt]{elsarticle}

\usepackage{amsmath}
\usepackage{amssymb}

\usepackage{graphicx}

\usepackage{doi}
\usepackage{hyperref}

\usepackage{xcolor}
\usepackage{longtable}
\usepackage{multirow}

\usepackage{enumitem}
\usepackage{float}

\biboptions{sort&compress}

\newdefinition{remark}{Remark}

\begin{document}

\begin{frontmatter}

\title{Group foliations, invariant solutions, and conservation laws of the geopotential forecast equation}

\author[address]{E.~I. Kaptsov\corref{correspondingauthor}}

\ead{evgkaptsov@math.sut.ac.th}


\affiliation[address]{organization={School of Mathematics, Institute of Science, Suranaree University of Technology},
postcode={30000},
city={Nakhon Ratchasima},
country={Thailand}}

\cortext[correspondingauthor]{Corresponding author}

\date{\today}

\begin{abstract}
Despite the large number of publications on symmetry analysis of the geopotential forecast equation, its group foliations laws have not been considered previously. The present publication aims to address this shortcoming. 
First, group foliations are constructed for the equation, and based on them, invariant solutions are derived, some of which generalize previously known exact solutions. There is also a discussion of the pros and cons of the group foliation approach. 
In addition, the rest of the paper is dedicated to a comprehensive discussion of conservation laws. All possible second-order conservation laws of the geopotential forecast equation are obtained through direct calculations, and a number of higher-order conservation laws are derived using the known symmetries of the equation.
\end{abstract}


\end{frontmatter}



\section{Introduction}

The geopotential forecast equation\footnote{Also known as barotropic vorticity equation on the beta-plane.} \cite{bk:Kibel1957} is used for  forecasting geopotential on regions of rotating air masses situated at intermediate altitudes within the atmosphere (`mid-level vortices'). These vortices are often associated with various weather phenomena and can play a significant role in the development of weather systems~\cite{bk:WallaceHobbs[2006]}. 
This equation is often written as
\begin{equation} \label{EqMain0}
\zeta_t - H_y \zeta_x + H_x (\zeta + f_0 + \beta y)_y = 0,
\end{equation}
where $\beta$ and $f_0$ are constants,
$(-H_y, H_x)$ is the two-dimensional velocity potential,
$\zeta = H_{xx} + H_{yy}$ represents relative vorticity, 
$f_0 + \beta y$ is the $\beta$-approximation of the Coriolis parameter~\cite{bk:WallaceHobbs[2006]},
and $\zeta + f_0 + \beta y$ is the absolute velocity.

\smallskip

When constructing exact solutions, reductions and conservation laws of equations of mathematical physics, methods of Group Analysis~\cite{bk:Ovsiannikov1978,bk:Olver[1986]} are typically among the main tools employed. 
Therefore, symmetries and invariant solutions of equation~(\ref{EqMain0}), as well as models closely related to it, have been extensively studied in numerous publications~\cite{
HUANG2004428,
bk:BihloPop_Geostr_InvSols[2011],
Bihlo_2009a,
Bihlo_2009b,
Fei_2004,
TangXiao-Yan_2008,
bk:Katkov_Geostr[1965],
bk:Katkov_Geostr[1966],
bk:HandbookLie_v2}.
The present work aims to address some gaps in previous research.

First, we consider the group foliations (stratifications) of equation~(\ref{EqMain0}). 
The construction of invariant solutions of equations is often simplified by utilizing their group foliations. 
There are other advantages of this approach, which we will discuss in the appropriate sections. 
Group foliations are usually constructed for equations that admit infinite-dimensional~Lie algebras. 
Many examples of foliations have been constructed for various equations of this kind~\cite{
bk:Ovsyannikov_strat[1970],
bk:Vereshchagina_strat[1974],
bk:Ovsiannikov1978,
bk:WinternitzMartina_strat[2001],
bk:NitkuSheftel_strat[2001],
bk:Golovin_strat[2003],
bk:Golovin_strat[2004inv],
bk:ChangZheng_start[2005],
bk:Pohjanpelto_strat[2010],
bk:DorKapMel_SW2D}.

Secondly, we thoroughly study the conservation laws of both second and higher orders possessed by equation~(\ref{EqMain0}).

\medskip

This paper is organized as follows. 
In Section~\ref{sec:sec1}, the Lie algebras of equation~(\ref{EqMain0}) are given for the cases $\beta\neq 0$ and $\beta=0$. 
Section~\ref{sec:sec2} is devoted to the construction of group foliations of the geopotential forecast equation. 
The theory necessary for these purposes is also provided. The foliations for the cases $\beta \neq 0$ and $\beta = 0$ are considered separately, and it turns out that the case $\beta=0$ is much more complicated. In Section~\ref{sec:sec3}, based on the constructed foliation for the case $\beta\neq 0$, invariant solutions are derived. Advantages and disadvantages of the method are discussed here as well. For illustrative purposes, one of the obtained solutions is considered numerically. 
Conservation laws of equation~(\ref{EqMain0}) are given in Section~\ref{sec:sec4}.
In Conclusion the obtained results are discussed.

\section{Lie algebras admitted by the geopotential forecast equation}
\label{sec:sec1}

Further we prefer to consider equation~(\ref{EqMain0}) in terms of the function $H$ and its derivatives in the following equivalent form~\cite{bk:HandbookLie_v2}
\begin{equation} \label{EqMain}
F = (H_{xx} + H_{yy})_t + H_x (H_{xx} + H_{yy})_y - H_y (H_{xx} + H_{yy})_x + \beta H_x = 0.
\end{equation}
The latter equation admits the following Lie algebra~\cite{bk:HandbookLie_v2}
\begin{equation} \label{alg}
\def\arraystretch{2}
\begin{array}{c}
X_1 = \partial_t,
\qquad
X_2 = \partial_y,
\qquad
X_3 = -t \partial_t + x \partial_x + y \partial_y + 3 H \partial_H,
\\
X_{\infty} = f \partial_x + (g - y f^\prime) \partial_H,
\end{array}
\end{equation}
where $f$ and $g$ are arbitrary smooth functions of~$t$. 
Further we assume $f \neq 0$ and $f^\prime = 0 \Rightarrow g \neq 0$, as otherwise there is no $\partial_x$ and $\partial_H$ generators which must be admitted by~(\ref{EqMain}).

The commutation relations for generators~(\ref{alg}) are
\[
\def\arraystretch{2}
\begin{array}{l}
    [X_1, X_2] = 0,
    \qquad
    [X_1, X_3] = - X_1,
    \qquad
    [X_2, X_3] = X_2,
    \\ {}
    [X_1, X_{\infty}] = f^{\prime} \partial_x + \left(g^{\prime} - y f^{\prime\prime}\right) \partial_H,
    \qquad
    [X_2, X_{\infty}] = -f^{\prime} \partial_H, 
    \\ {}
    [X_3, X_{\infty}] = -(f + t f^{\prime}) \partial_x - \left(t g^{\prime} + 3 g  - y (t f^{\prime\prime} + 2 f^{\prime}) \right) \partial_H.
\end{array}
\]
One sees that the generators $X_1$, $X_2$, and $X_3$ form a three-dimensional subalgebra of the infinite-dimensional Lie algebra~(\ref{alg}).

\medskip

In case $\beta = 0$, the admitted Lie algebra consists of the generators~\cite{Bihlo_2009b}
\begin{equation} \label{alg0}
\def\arraystretch{2}
\begin{array}{c}
X_1^0 = \partial_t, 
\qquad
X_2^0 = t \partial_t - H \partial_H,
\qquad
X_3^0 = y \partial_x - x \partial_y,
\qquad
X_4^0 = x \partial_x + y \partial_y + 2 H \partial_H,
\\
\displaystyle
X_5^0 = t y \partial_x - t x \partial_y - \frac{x^2 + y^2}{2} \,\partial_H,
\qquad
X_{\infty}^0 = \varphi \partial_x + \psi \partial_y + (x \psi^\prime - y \varphi^\prime + \chi) \partial_H,
\end{array}
\end{equation}
where $\varphi$, $\psi$, and $\chi$ are arbitrary smooth functions of~$t$. It is assumed that $\varphi \psi \chi \neq 0$, otherwise the symmetries $\partial_x$, $\partial_y$, or $\partial_H$, which must be admitted by~(\ref{EqMain}), are not in the algebra.

\begin{remark}
In case $\beta \neq 0$, one can assume $\beta=1$ by means of the equivalence transformation
\[
x = \widetilde{x},
\qquad
y = \widetilde{y},
\qquad
t = \widetilde{t} / \beta,
\qquad
H = \beta \widetilde{H}.
\]
Recall that equivalence transformations admitted by an equation do not affect its group structure~\cite{bk:Ovsiannikov1978}. 
Nevertheless, we do not use the specified transformation here, as it is convenient for further discussion to preserve the parameter~$\beta$.
It is also worth noting that sufficiently large values of~$\beta$ contribute to the formation of Rossby waves~\cite{bk:WallaceHobbs[2006]}.
\end{remark}

\section{Group foliations of equation~(\ref{EqMain})}
\label{sec:sec2}

\subsection{Preliminary discussion and basic theory}

The construction of a group foliation is based on the theory of differential invariants~\cite{bk:Ovsiannikov1978} (a brief and clear presentation of the main ideas can be found, for example, in~\cite{bk:Ovsyannikov_strat[1970],bk:Golovin_strat[2004inv]}) and allows one to move from the original equation (or a system) to two systems of equations: automorphic and resolving. These systems are written in terms of differential invariants of the generator~$X$ with respect to which the foliation is carried out, as well as new functions depending on the differential invariants as new independent variables~(usually their number is equal to the number of independent variables in the original equation). The resolving system is obtained as a set of compatibility conditions for the original equation and the automorphic system and includes only new functions and new dependent variables. Each particular solution of the resolving system allows one to fix a particular form of the automorphic system dependent on it. All solutions of such a fixed automorphic system are related by the transformations corresponding to~$X$, so for its complete integration it is enough to know any of its particular solutions. To obtain an automorphic system with the widest possible set of solutions, it is natural to choose as~$X$ a generator corresponding to an infinite-dimensional group of transformations. In this case, the resolving system will admit a finite-dimensional Lie algebra isomorphic to a remaining finite-dimensional `part' of the Lie algebra admitted by the original equation. This often simplifies the further work with optimal systems of subalgebras and invariant solutions.

\medskip


As mentioned above, a detailed theory of differential invariants and group foliations can be found in~\cite{bk:Ovsiannikov1978}, while here we only briefly outline the theory required to construct group foliations.

\smallskip

Consider an equation given on the space of variables 
\[
\underset{k}{\mathcal{Z}} = \mathcal{X} \times \mathcal{U} \times \underset{1}{\mathcal{U}} \times \cdots \times \underset{k}{\mathcal{U}},
\]
where $\mathcal{X}=(x^1, \dots, x^m)$ and $\mathcal{U} = (u^1, \dots, u^n)$ are independent and dependent variables (in our case, $\mathcal{X} = (x_1, x_2, x_3) = (t,x,y)$, $\mathcal{U} = (H)$), and $\underset{i}{\mathcal{U}}$ denotes the prolongation of the dependent variables~$\mathcal{U}$ to the~$i$-th order derivatives of~$\mathcal{U}$ with respect to~$\mathcal{X}$.

Having chosen the generator
\[
    X = \underset{0}{X} = \xi^i \partial_{x^i} + \eta^j \partial_{u^j},
\]
for which the foliation is carried out, we consider its prolongations $\underset{k}{X}$, $k=1, 2, \dots$, given by the standard prolongation formulas~\cite{bk:Ovsiannikov1978,bk:Olver[1986]}:
\[
\underset{k}{X} = \underset{0}{X}
+ \zeta^\alpha \partial_{u^\alpha} + \cdots
+ \zeta^\alpha_{i_1 , \dots, i_k} \partial_{u^\alpha_{i_1 , \dots, i_k}},
\qquad
\zeta^\alpha_{i_1 , \dots, i_s} = D_{i_1} \cdots D_{i_s}(\eta^\alpha - u^\alpha_j \xi^j) + \xi^j u^\alpha_{j i_1, \dots, i_s},
\]
where $D_{i}$ is the operator of total differentiation with respect to~$x_i$. 
Here and further we use summation over repeating indices.

Splitting each of the prolongations with respect to the arbitrary functions included in the coefficients of the generator, for~$k=0, 1, 2, \dots$ one obtains 
sets of generators
\[
\underset{k}{X}^\mu = 
\xi^{\mu, i} \partial_{x^i} + \eta^{\mu, j} \partial_{u^j}
+ \zeta^{\mu, \alpha} \partial_{u^\alpha} + \cdots
+ \zeta^{\mu, \alpha}_{i_1 , \dots, i_k} \partial_{u^\alpha_{i_1 , \dots, i_k}},
\]
for which the following general rank numbers are considered
\[
\displaystyle
r_k = \max \; \textrm{rank} \; 
\begin{pmatrix}
\xi^{1,1}   & \cdots & \xi^{1,m} & \eta^{1, 1} & \cdots & \eta^{1, n} & \zeta^{1, \alpha} & \cdots & \zeta^{1, \alpha}_{i_1 , \dots, i_k}
\\
\vdots &  & &  & \ddots &  &  &  & \vdots
\\
\xi^{\mu,1}   & \cdots & \xi^{\mu,m} & \eta^{\mu, 1} & \cdots & \eta^{\mu, n} & \zeta^{\mu, \alpha} & \cdots & \zeta^{\mu, \alpha}_{i_1 , \dots, i_k}
\end{pmatrix}.
\]
For further analysis it is also convenient to introduce the numbers
\[
\nu_k = \dim \underset{k}{\mathcal{Z}} = m + n \, \binom{k + m}{m}.
\]
In our case, $m = 3$, $n = 1$, and thus
\[
\nu_0 = 4,
\quad
\nu_1 = 7,
\quad
\nu_2 = 13,
\quad
\nu_3 = 23,
\quad
\text{etc.}
\]
The number of functionally independent differential invariants of~$X$ of order $\leqslant k$ is given by
\[
\rho_k = \nu_k - r_k.
\]
To construct a group foliation, $m$ new independent variables are usually selected from among the differential invariants, and $\rho_k - m$ variables become new dependent ones. It is required that $\rho_k > m$, otherwise it is not possible to construct a foliation.
Equations connecting new independent variables and dependent differential invariants form an automorphic system, while the conditions of 
their compatibility with the original equation give a resolving system.

Differential invariants can be found either directly or using invariant differentiation operators. It is known that for any Lie group there are exactly~$m$ (according to the number of independent variables) operators of invariant differentiation and a finite number of basis differential invariants, with the help of which all other differential invariants can be derived. Here we do not consider in detail the operators of invariant differentiation and the basis of differential invariants and refer the interested reader to~\cite{bk:Ovsiannikov1978}.

\medskip

In what follows we separately consider group foliations for two cases, $\beta \neq 0$ and $\beta = 0$, with respect to the generators $X_\infty$ and $X^0_\infty$.

\subsection{Case $\beta \neq 0$}

Here we consider the group foliation corresponding to the generator $X_{\infty}$.

Splitting the $k$-th ($k = 0, 1, 2, \dots$) prolongations of the generator $X_{\infty}$ with respect to the arbitrary functions~$f$,~$g$ and their derivatives, 
one derives the following sets $S_i$ of generators.
\begin{enumerate}[label=\arabic*)]
  \item 
  For $k=0$:
  \[
  S_0 = \{ \partial_x, \partial_H, y \partial_H \}.
  \]
  Then, we find $r_0 = 2$, and there are $\rho_0 = \nu_0 - r_0 = 2$ independent differential invariants: $t$ and $y$.
  
  \item
  For $k=1$:
  \[
  S_1 = S_0 \cup \{ \partial_{H_t}, y \partial_{H_t}, y \partial_{H} + H_x \partial_{H_t} + \partial_{H_y} \},
  \qquad 
  r_1 = 4,
  \]
  and there are $\nu_1 - r_1 = 3$ differential invariants of order~$\leqslant 1$:
  \[ t, \quad y, \quad H_x. \]
  
  \item
  For $k=2$, the set $S_2$ is
  \[
  \displaystyle
  S_1 \cup \{ \partial_{H_{tt}}, y \partial_{H_{tt}}, y \partial_{H_{t}} + H_x \partial_{H_{tt}} + \partial_{H_{ty}},
                        y \partial_{H} + \partial_{H_{y}} + H_x \partial_{H_{t}} + H_{xy} \partial_{H_{ty}} + 2 H_{tx} \partial_{H_{tt}}
                        + H_{xx} \partial_{H_{tx}} \}.
  \]
  The corresponding rank $r_2 = 6$, and there are $\nu_2 - r_2 = 7$ independent differential invariants of the order~$\leqslant 2$:
  \[
  t, \quad y, \quad H_x, \quad H_{xx}, \quad H_{xy}, \quad H_{yy}, \quad H_y H_{xx} - H_{tx}.
  \]
  
  \item
  For $k=3$, the set $S_3$ is quite cumbersome and thus it is not given here. In this case the rank $r_3 = 8$.
  There are $\nu_3 - r_3 = 15$ independent differential invariants of the order~$\leqslant 3$:
  \begin{equation}\label{invsOrder3}
  \def\arraystretch{2}
  \begin{array}{c}
    t, \quad y, \quad H_x, \quad H_{xx}, \quad H_{xy}, \quad H_{yy}, \quad H_y H_{xx} - H_{tx}, \\
    H_{xxx},\quad H_{yyy},\quad H_{xxy}, \quad H_{xyy}, \quad H_{y} H_{xxx}-H_{txx}, \quad H_{y} H_{xyy}-H_{tyy}, \\
    H_{y} H_{xxy}-H_{tx y}, \quad H_{xxx} H_{y}^2+(H_{xx} H_{xy}-2 H_{txx}) H_{y}-H_{ty} H_{xx}+H_{ttx}.
  \end{array}
  \end{equation}
\end{enumerate}
Equation~(\ref{EqMain}) can be expressed in terms of invariants~(\ref{invsOrder3}). 
Actually, 11 out of 15 invariants~(\ref{invsOrder3}) can be obtained from four basis invariants 
\begin{equation} \label{binvs}
b_{0,1} = t,
\qquad
b_{0,2} = y,
\qquad
b_1 = H_x,
\qquad
b_2 = H_y H_{xx} - H_{tx},
\end{equation}
applying to them the following operators of invariant differentiation
\begin{equation} \label{deltas}
\delta_1 = D_t + \frac{x f^\prime}{f} D_x,
\qquad
\delta_2 = D_x,
\qquad
\delta_3 = D_y.
\end{equation}
In terms of basis invariants (\ref{binvs}) and operators (\ref{deltas}), equation (\ref{EqMain}) can be rewritten as
\[
(b_4 + \beta) b_1 + b_5 = 0,
\]
where $b_4$ and $b_5$ are differential invariants given by
\[
\def\arraystretch{2}
\begin{array}{l}
b_4 = \delta_2 \delta_3 b_1 +\delta_3 b_3 = H_{xxy} + H_{yyy}, 
\\
\displaystyle
b_5 = \left( 2 {b_3} \delta_2 \delta_3 b_1 + \delta_3 b_3 \, \delta_2 b_1 -\delta_3^2 b_2 - \delta_1 \delta_2 b_3 \right) \frac{\delta_3^2 b_1}{\delta_2^2 b_3} + \delta_3 b_1 \, \delta_2 b_1+\delta_1 b_3 -\delta_2 b_2 \\
\qquad\qquad\qquad = H_{txx} + H_{tyy} - (H_{xxx} + H_{xyy}) H_{y}.
\end{array}
\]
These invariants are also depend on the invariant
\[
b_3 = b_3(b_1, b_2) = \frac{\delta_1 \delta_3 b_1 + \delta_3 b_2}{\delta_2 b_1} + \frac{(\delta_2 b_1 \delta_3 b_1 -\delta_1 \delta_2 b_1 -\delta_2 b_2) \delta_2 \delta_3 b_1} {\delta_2 b_1 \delta_2^2 b_1} = H_{yy}.
\]
Here it worth to provide the commutators of the operators~(\ref{deltas})
\[
[\delta_1, \delta_2] = -\frac{f^\prime}{f} \delta_2,
\qquad
[\delta_1, \delta_3] = 0,
\qquad
[\delta_2, \delta_3] = 0,
\]
as $\delta_1$ and $\delta_2$ do not commute, so their order in the given representation is important.

\bigskip

To construct a group foliation for~(\ref{EqMain}), we choose $m=3$ 
new independent variables, $t$, $y$, and $h = H_x$, and assume
\begin{equation} \label{sysAE}
H_y H_{xx} - H_{tx} = U(t, y, h), \quad
H_{xx} = V(t, y, h), \quad
H_{xy} = W(t, y, h), \quad
H_{yy} = Z(t, y, h),
\end{equation}
where $U$, $V$, $W$, and $Z$ are some smooth enough functions of the chosen independent variables.

To obtain compatibility conditions of (\ref{sysAE}) and (\ref{EqMain}), it is necessary to differentiate these equations several times with respect to~$t$, $x$, and~$y$, and eliminate from the resulting system of equations all derivatives of the variable~$H$, except~$H_x = h$. 
Finally, we arrive at the following overdetermined quasi-linear system.
\begin{equation} \label{sysRE}
\def\arraystretch{2}
\begin{array}{c}
    {V_y}+ {W} {V_h} - {V} {W_h} = 0,
    \qquad
    {W_y}+ {W} {W_h}- {V} {Z_h} = 0,
    \\
    {V_t}+{V}{U_h} - {U}{V_h} -{V} {W} = 0,
    \qquad
    {W_t}+{U_y}+{W}{U_h}- {U} {W_h}-{V} {Z} = 0,
    \\
    Z_t - U Z_h - V U_h + \left(W Z_h+V W_h+Z_y+\beta\right) h + V W = 0.
\end{array}
\end{equation}
Thus, we have split equation (\ref{EqMain}) into the automorphic system~(\ref{sysAE}) and the resolving system~(\ref{sysRE}).

\medskip

Standard calculations show that system (\ref{sysRE}) admits the following generators
\begin{equation} \label{algY}
\displaystyle
Y_1 = \partial_{t},
\qquad
Y_2 = \partial_{y},
\qquad
Y_3 = -t\partial_{t} + y \partial_{y} + 2 h \partial_{h} 
        + 3 U \partial_{U}
        + V \partial_{V}
        + W \partial_{W}
        + Z \partial_{Z},
\end{equation}
which constitute a three-dimensional Lie algebra isomorphic to the subalgebra $\{ X_1, X_2, X_3 \}$ of~(\ref{alg}).

\begin{remark}
We study compatibility conditions of system~(\ref{sysRE}) using the Cartan criterion~\cite{bk:Pommaret} (one can find a more practical approach and additional references, e.g., in~\cite{bk:Meleshko[2005]}).
There are~5 equations and~12 derivatives of the functions $U$, $V$, $W$, and $Z$ with respect to $t$, $y$, and $h$. 
The rank of the Jacobian of the system with respect to these derivatives equals~5, thus the auxiliary constant
$\tau_0 = 12 - 5 = 7$. 

Next, we augment the system with the equations
\[
U_t = 0, \qquad V_t = 0, \qquad W_t = 0, \qquad Z_t = 0,
\]
which extends the original system to the system of~9 equations. The matrix of the coefficients of the derivatives is of rank~9. 
This leads to the second auxiliary constant $\tau_1 = 12 - 9 = 3$. Then, we extend the latter system with the equations
\[
U_ y= 0, \qquad V_y = 0, \qquad W_y = 0, \qquad Z_y = 0,
\]
which leads to a matrix of rank~$12$, and $\tau_2 = 12 - 12 = 0$. 

Now we can calculate the Cartan characters $\sigma_i = \tau_i - \tau_{i+1}$ and the Cartan number $Q$:
\[
\sigma_1 = \tau_0 - \tau_1 = 4,
\qquad
\sigma_2 = \tau_1 - \tau_2 = 3,
\qquad
Q = \tau_0 + \tau_1 = 10.
\]
Finally, we differentiate the equations of system~(\ref{sysRE}) with respect to $t$, $y$, and~$h$ to derive a prolonged system of~15 equations of the second order which involve up to~$4 C^2_3 = 24$ second-order derivatives of the functions $U$, $V$, $W$, and~$Z$.
Rank of the Jacobian of this system with respect to the second order derivatives equals~14.
Thus, there are~$Q_1 = 24 - 14 = 10$ `free' second-order derivatives in the prolonged system. As~$Q_1=Q$, the Cartan criterion is satisfied.
This means that system~(\ref{sysRE}) is consistent and in involution. 
Arbitrariness of its general solution is determined by the Cartan characters~$\sigma_1$ and~$\sigma_2$:~3 functions of~2 arguments and~4 functions of~1 argument.
\end{remark}

\subsection{Case $\beta = 0$}

Consider the representation of~(\ref{EqMain}) in terms of differential invariants of the generator~$X^0_{\infty}$. 

The operators of invariant differentiation corresponding to the generator $X^0_{\infty}$ are
\[
\delta_1^0 = D_t + \frac{x {\varphi}^\prime}{{\varphi}} \, D_x + \frac{x {\psi}^\prime}{{\varphi}} \, D_y,
\qquad
\delta_2^0 = \left( H_y + \frac{x {\varphi}^\prime}{{\varphi}} \right) \, D_x,
\qquad
\delta_3^0 = \left( H_x - \frac{x {\psi}^\prime}{{\varphi}} \right) \, D_y,
\]
and the basis of differential invariants of the generator $X^0_{\infty}$ is the following.
\begin{equation} \label{binvs0}
b^0_{0} = t, 
\qquad
b^0_{1} = H_{xx},
\qquad
b^0_{2} = H_{xy},
\qquad
b^0_{3} = H_{yy}.
\end{equation}
In terms of basis invariants (\ref{binvs0}) and operators (\ref{deltas}), equation (\ref{EqMain}) (for~$\beta = 0$) can be rewritten as
\[
(\delta^0_1 - \delta^0_2 + \delta^0_3) b^0_1 + (\delta^0_1 + \delta^0_3) b^0_3 - \frac{\delta^0_2 b^0_2 \, \delta^0_3 b^0_2}{\delta^0_3 b^0_1} = 0.
\]

In case $\beta=0$, it turns out that there is no differential invariants of order~1 for the generator~$X^0_{\infty}$, which entails a significant complication in the construction of a group foliation. To construct an automorphic system, one introduces a dependence of the invariant~$H_{yy}$ as well as~7 third-order invariants of the generator~$X^0_{\infty}$, namely,
\[
\def\arraystretch{1.5}
\begin{array}{c}
H_{xxx}, \quad 
H_{xxy}, \quad 
H_{yyy}, \quad
H_{xyy}, \quad
H_{txx} + H_{x} H_{xxy}-H_{y} H_{xxx},
\\
H_{tyy} + H_{x} H_{yyy}-H_{y} H_{xyy}, \qquad 
H_{txy} + H_{x} H_{xyy}-H_{y} H_{xxy},
\end{array}
\]
on the new independent variables~$t$, $H_{xx}$, and~$H_{xy}$. 
The compatibility conditions of these equations and the original system lead to a resolving system of~11 equations, 4 of which are non-linear and quite cumbersome, thus we do not present that system here. Despite the fact that the resolving system is often simpler than the original equation, 
obtaining solutions or at least analyzing the resolving system in this case is quite problematic and does not seem to be productive.
Thus, further we stay focused only on the group foliation~(\ref{sysAE}),~(\ref{sysRE}) constructed for the case~$\beta \neq 0$.

\section{Invariant solutions in context of group foliation~(\ref{sysAE}),~(\ref{sysRE})}
\label{sec:sec3}

The group foliation method has at least three advantages over the standard method of constructing invariant solutions. 
Firstly, group foliations allow one to `discard' an infinite-dimensional part of a Lie algebra and consider optimal systems of subalgebras for a finite-dimensional part, which allows one to significantly simplify further analysis.
Secondly, the method of group foliations sometimes extends the classes of invariant solutions derived for the same subalgebras by a standard procedure~\cite{bk:Golovin_strat[2003]}. We will discuss this in more detail later in this section.

The third advantage is that any particular solution of an automorphic system allows one to obtain the entire set of solutions of this system using the generator with respect to which the foliation has been constructed. 
For example, one can verify that a particular solution of the resolving system (\ref{sysRE}) is
\[
\def\arraystretch{2}
\begin{array}{c}
\displaystyle
U = \left(\frac{\beta {\kappa}}{{\kappa}^2 + {\nu}^2}  - {\nu} h\right) \sqrt{{\rho}^2 {\kappa}^2 - h^2},
\qquad
V = -{\kappa} \sqrt{{\rho}^2 {\kappa}^2 - h^2},
\\
\displaystyle
W = -{\nu} \sqrt{{\rho}^2 {\kappa}^2 - h^2},
\qquad
Z = -\frac{{\nu}^2}{{\kappa}} \sqrt{{\rho}^2 {\kappa}^2 - h^2},
\end{array}
\]
where ${\kappa}$, ${\nu}$, and ${\rho}$ are some constants such that $({\kappa}^2 + {\nu}^2) \kappa \neq 0$. 
Then, the automorphic system~(\ref{sysAE}) is brought to the form
\begin{equation}
\def\arraystretch{2}
\begin{array}{c} \label{sysAEEg1}
\displaystyle
H_{xx} = -{\kappa} \sqrt{{\rho}^2 {\kappa}^2 - H_{x}^2}, \quad
H_{xy} = -{\nu} \sqrt{{\rho}^2 {\kappa}^2 - H_{x}^2}, \quad
H_{yy} = -\frac{{\nu}^2}{{\kappa}} \sqrt{{\rho}^2 {\kappa}^2 - H_{x}^2}, \\
\displaystyle
H_{tx} - H_y H_{xx} = \left({\nu} {H_{x}} - \frac{\beta {\kappa}}{{\kappa}^2 + {\nu}^2}\right) \sqrt{{\rho}^2 {\kappa}^2 - H_{x}^2}.
\end{array}
\end{equation}
One can find the following particular solution of the later system
\begin{equation} \label{GaurvitzSol}
H_0 = {\rho} \cos({\kappa} (x - {\lambda} t) + {\nu} y) + {\mu} y,
\end{equation}
where ${\mu}$ and ${\lambda}$ are constants satisfying
\[
({\kappa}^2 + {\nu}^2)({\lambda} + {\mu}) = -\beta.
\]
This is ``Gaurvitz'' invariant solution~\cite{bk:Kibel1957}, \cite[p.~224]{bk:HandbookLie_v2} corresponding to the generator
$\partial_t + {\lambda} \partial_x$
admitted by~(\ref{EqMain}).  

Recall that system~(\ref{sysAEEg1}) is invariant with respect to the generator~$X_{\infty}$ which corresponds to the group of transformations 
\begin{equation} \label{XinfGrTransform}
T_{\infty}^{\varepsilon, f, g}: \qquad
\bar{t} = t, 
\qquad
\bar{x} = x + \varepsilon f,
\qquad
\bar{y} = y,
\qquad
\bar{H} = H + (g - y f^\prime) \varepsilon, 
\end{equation}
where $\varepsilon$ is a real parameter and the functions $f$ and $g$ are arbitrary smooth enough functions of~$t$. Thus,
\[
\begin{array}{l}
T_{\infty}^{\varepsilon, f_1, g_1} H_0 = {\rho} \cos({\kappa} (x - {\lambda} t - \varepsilon f_1) + {\nu} y) + (g_1 - y f_1^\prime) \varepsilon + {\mu} y, 
\\
\displaystyle
\qquad \ldots
\\
\displaystyle
(T_{\infty}^{\varepsilon, f_1, g_1} \circ \cdots \circ T_{\infty}^{\varepsilon, f_n, g_n}) H_0 
    = {\rho} \cos({\kappa} (x - {\lambda} t - \sum_{i=1}^n \varepsilon_i f_i) + {\nu} y) + \sum_{i=1}^n (g_i - y f_i^\prime) \varepsilon_i + {\mu} y,
\end{array}
\]
are again solutions of~(\ref{sysAEEg1}). 
Here we note that consecutive application of~(\ref{XinfGrTransform}) even to a simple particular solution can lead to quite  
fancy solutions of the same equation, which can be seen, for example, in~\cite{bk:xu2008algebraic}.

What is important, is that the \emph{entire} set of solutions of the automorphic system~(\ref{sysAEEg1}) can be generated by transformations~(\ref{XinfGrTransform}) applied to any particular solution~(say,~(\ref{GaurvitzSol})).

\smallskip

Although the role of system~(\ref{sysAEEg1}) by itself is not very clear from the point of view of practical applications, 
there are some cases where automorphic and resolving systems have a known physical meaning. Here it is worth mentioning a connection of such kind between the two-dimensional shallow water equations in Lagrangian and Eulerian coordinates that was revealed in~\cite{bk:DorKapMel_SW2D}:
the equations for the transition from Eulerian to Lagrangian coordinates turn out to be an automorphic system with respect to the relabeling symmetry, whereas the shallow water equations in Eulerian coordinates define the resolving system.\footnote{To the best of the author's knowledge, this fact was first established by Prof.~S.V.~Meleshko. Apparently, this result is also the case for many other hydrodynamic-type equations in Lagrangian coordinates.}
In this case, solutions of the automorphic system give trajectories of the particles of the medium.

\bigskip

Consider the autonomous system~(\ref{sysAE}). 
To obtain sets of dissimilar solutions of~(\ref{sysAE}) (and~(\ref{EqMain})), one finds optimal systems of subalgebras of algebra~(\ref{algY}).
According to~\cite{bk:PateraWinternitz1977}, this algebra corresponds to the pseudo-Euclidean group~$E(1,1)$ and its optimal systems of subalgebras are
\[
\def\arraystretch{1.5}
\begin{array}{ll}
  \dim 3: & \{ Y_1, Y_2, Y_3 \}, \\
  \dim 2: & \{ Y_1, Y_2 \},  \; \{ Y_1, Y_3 \}, \; \{ Y_2, Y_3 \}, \\
  \dim 1: & \{ Y_1 \}, \; \{ Y_2 \}, \; \{ Y_3 \}, \; \{ Y_1 \pm Y_2 \}.
\end{array}
\]
Further we consider the listed subalgebras and corresponding invariant solutions.

\subsection{The three-dimensional subalgebra $\{ Y_1, Y_2, Y_3 \}$}

In this case system~(\ref{sysRE}) has only a trivial solution
\[
U = V = W = 0, \qquad Z = z_0 = \textrm{const},
\]
and $\beta h = \beta H_x = 0$, which leads to
\begin{equation} \label{H_A3}
H = \frac{z_0 y^2}{2} + {\tau}_1 y + {\tau}_2,
\end{equation}
where ${\tau}_1$ and ${\tau}_2$ are arbitrary functions of~$t$.

\begin{remark}
Solution (\ref{H_A3}) is equivalent to the solution
\[
\bar{H} = \frac{z_0 y^2}{2}
\]
by means of transformation~(\ref{XinfGrTransform}) with $\varepsilon g=-{\tau}_2$ and $\varepsilon f^{\prime} = {\tau}_1$. 
Thus, in what follows, we will, if possible, omit one or two of these terms of~(\ref{H_A3}).
But this should be done carefully as transformation~(\ref{XinfGrTransform}) also affects the variable~$x$.
\end{remark}

\subsection{Two-dimensional subalgebras}

For the two-dimensional subalgebras of~(\ref{algY}), system~(\ref{sysRE}) reduces to systems of ordinary differential equations. 
Although in most cases these systems cannot be solved analytically, numerical solutions can be found for them using standard methods. 
Further we provide some particular analytical solutions.

Since finding a general solution of even a reduced system of resolving equations is problematic, here and further we adhere to the following approach. 
As a rule, the first two equations of system~(\ref{sysRE}) after reduction on a subgroup take on a fairly simple form and some solutions can be found for them. 
We choose these solutions as ansatzes for the remaining equations of the system. 
From the resulting solutions obtained in this way, we select only those that provide independent solutions to the original system~(\ref{EqMain}).

\subsubsection{Subalgebra $\{ Y_1, Y_2 \}$}

\medskip

In this case the functions $U$, $V$, $W$, and $Z$ depend only on $h$.
System (\ref{sysRE}) reduces to
\begin{equation}
\def\arraystretch{1.75} \label{RErdcA21}
\begin{array}{c}
V W^{\prime}  - W V^{\prime} = 0, \quad
V Z^{\prime} - W W^{\prime} = 0, \\
U V^{\prime} - V U^{\prime} + V W = 0, \quad
U W^{\prime} - W U^{\prime} + V Z = 0,
\\
V U^{\prime} + U Z^{\prime} - (V W^{\prime} + W Z^{\prime} + \beta) h - V W = 0.
\end{array}
\end{equation}

We consider the following particular solutions of the reduced system~(\ref{RErdcA21}).
\begin{itemize}
  \item
  Solution
  \[
  U = \frac{\beta h}{Z^\prime}, \qquad V = W = 0, \qquad Z = Z(h)
  \]
  of system~(\ref{RErdcA21}) leads to the invariant solution
  \begin{equation} \label{A21isol1}
  H = A (y^2 + 2 x) e^{-\beta t} + {\tau} y,
  \end{equation}
  where ${\tau}$ is an arbitrary function of $t$, and $A$ is a constant.
  \item
  Solution
  \[
  U = \frac{1}{C_1}, \qquad 
  V = 0, \qquad 
  W = C_2, \qquad 
  Z =  C_3 - \frac{\beta h}{C_2} - \frac{\beta \ln(C_1 C_2 h - 1)}{C_1 C_2^2}
  \]
  leads to
  \begin{equation} \label{A21isol2}
  \displaystyle
  H = \left(
        \frac{A_3^6 \zeta^2}{4}
        +A_3^4 y \zeta
        -\frac{A_3^2 y^2}{2} \left(\zeta + 2\right)
        +\frac{y^3}{3}
    \right) \beta
    -\frac{1}{2}{A_3^6 \beta \zeta^2 \ln{\zeta}}
    + A_1 A_3 x (\zeta + 1)
    + A_2 y^2 
    + y {\tau},
  \end{equation}
  where ${\tau}$ is an are arbitrary function of~$t$; $A_1$, $A_2$, and $A_3$ are constants, and
  \[
  \zeta = \zeta(t, y) = C_1 C_2^2 y+(C_1 C_4-t) C_2-1,
  \qquad
  A_1^2 = \frac{1}{C_1},
  \qquad
  A_2 = \frac{C_3}{2},
  \qquad
  A_3^2 = \frac{1}{C_1 C_2^2}.
  \]
  
  \item
  Solution
  \[
  \def\arraystretch{2}
  \begin{array}{c}
    \displaystyle
    U = \left( C_3 + \frac{C_1 \beta h}{C_1^2 + C_3^2} \right) \sqrt{C_2 - h^2}, \qquad 
    V = C_1 \sqrt{C_2 - h^2}, 
    \\
    \displaystyle
    W = C_3 \sqrt{C_2 - h^2}, \qquad 
    Z = \frac{C_3^2}{C_1} \sqrt{C_2 - h^2}
  \end{array}
  \]
  can be easily brought to system~(\ref{sysAEEg1}), which was discussed in the beginning of the section. 
  Then, one derives an invariant solution of type~(\ref{GaurvitzSol}).
\end{itemize}

Obviously, any linear combination\footnote{Actually, with some simple restrictions on their constant coefficients.} of solutions of the form (\ref{GaurvitzSol}), (\ref{A21isol1}), and~(\ref{A21isol2}) 
is also an invariant solution of~(\ref{EqMain}).

\subsubsection{Subalgebra $\{ Y_1, Y_3 \}$}

\medskip

In this case we consider
\[
U = y^3 \widetilde{U}(\lambda), \qquad
V = y \widetilde{V}(\lambda), \qquad
W = y \widetilde{W}(\lambda), \qquad
Z = y \widetilde{Z}(\lambda), \qquad
\lambda = h y^{-2},
\]
and system (\ref{sysRE}) reduces to
\begin{equation}
\def\arraystretch{1.75} \label{RErdcA22}
\begin{array}{c}
(\widetilde{W}-2 \lambda) \widetilde{V}^{\prime}-\widetilde{V} \widetilde{W}^{\prime}+\widetilde{V} = 0,
\qquad
(\widetilde{W}-2 \lambda) \widetilde{W}^{\prime}-\widetilde{V} \widetilde{Z}^{\prime}+\widetilde{W} = 0,
\\
\widetilde{V} \widetilde{U}^{\prime}-\widetilde{V}^{\prime} \widetilde{U}-\widetilde{V} \widetilde{W}  = 0,
\qquad
(\widetilde{W}-2 \lambda) \widetilde{U}^{\prime}-\widetilde{U} \widetilde{W}^{\prime}-\widetilde{V} \widetilde{Z}+3 \widetilde{U} = 0,
\\
\widetilde{V} \widetilde{U}^{\prime}-\lambda \widetilde{V} \widetilde{W}^{\prime}-(\widetilde{W} \lambda-2 \lambda^2-\widetilde{U}) \widetilde{Z}^{\prime}-\widetilde{V} \widetilde{W}-\lambda (\beta+\widetilde{Z}) = 0.
\end{array}
\end{equation}
Some particular solutions of the latter system can be found for specific forms of the function~$\widetilde{W}$.
\begin{itemize}
\item
If ${\widetilde{W}} = w_0^2 = \textrm{const}$, the latter system has the solution
\[
\widetilde{U} = w_0^2 (2 \lambda - w_0^2) + \frac{w_0 \beta}{2}\,\sqrt{w_0^2-2 \lambda}, 
\qquad
\widetilde{V} = -w_0 \sqrt{w_0^2-2 \lambda}, 
\qquad
\widetilde{W} = w_0^2,
\qquad
\widetilde{Z} = \frac{w_0^3 - 2 w_0 \lambda}{\sqrt{w_0^2-2 \lambda}} - \beta.
\]
This leads to the following invariant solution
\begin{equation} \label{A22isol1}
\displaystyle
H = \frac{w_0^2}{2} (y^2 - x^2) {\tau}
    +\frac{w_0^2}{6} (3 y^2 - x^2) x
    -\frac{w_0^2}{2} x {\tau}^2
    - \frac{\beta y^3}{6}
    + y {\tau}^{\prime},
\end{equation}
where ${\tau}$ is an arbitrary functions of~$t$.

\item
Assuming ${\widetilde{W}} = \lambda$, one derives the following solution of~(\ref{RErdcA22})
\[
\widetilde{U} = \frac{\lambda^2 - C_1 (\beta + C_1)}{2}, 
\qquad
\widetilde{V} = C_1, 
\qquad
\widetilde{W} = \lambda,
\qquad
\widetilde{Z} = -\beta - C_1,
\]
where $C_1$ is constant. This leads to
\begin{equation} \label{A22isol2}
H = {\tau}_1 x y 
    + \frac{C_1}{2} x^2 y
    - \frac{\beta + C_1}{6}  y^3 + y {\tau}_2,
\end{equation}
where ${\tau}_1$ and ${\tau}_2$ are some functions of~$t$ satisfying
\[
{\tau}_1^{\prime\prime} + \frac{{\tau}_1^2}{2} - C_1 {\tau}_2 = 0.
\]

\end{itemize}

\subsubsection{Subalgebra $\{ Y_2, Y_3 \}$}

\medskip

In this case,
\[
U = t^{-3} \widetilde{U}(\lambda), \qquad
V = t^{-1} \widetilde{V}(\lambda), \qquad
W = t^{-1} \widetilde{W}(\lambda), \qquad
Z = t^{-1} \widetilde{Z}(\lambda), \qquad
\lambda = t^2 h,
\]
and system (\ref{sysRE}) reduces to
\[
\def\arraystretch{1.75}
\begin{array}{c}
\widetilde{W} \widetilde{V}^{\prime}-\widetilde{V} \widetilde{W}^{\prime} = 0,
\qquad
\widetilde{W} \widetilde{W}^{\prime} -\widetilde{V} \widetilde{Z}^{\prime} = 0,
\\
\widetilde{V} \widetilde{U}^{\prime}+(2 \lambda-\widetilde{U}) \widetilde{V}^{\prime}-(\widetilde{W}+1) \widetilde{V}  = 0,
\qquad
\widetilde{W} \widetilde{U}^{\prime}+(2 \lambda-\widetilde{U}) \widetilde{W}^{\prime}-\widetilde{W}-\widetilde{V} \widetilde{Z} = 0,
\\
\lambda \widetilde{V} \widetilde{W}^{\prime}-\widetilde{V} \widetilde{U}^{\prime}+((\widetilde{W} +2) \lambda-\widetilde{U}) \widetilde{Z}^{\prime}+\widetilde{W} \widetilde{V}-\widetilde{Z}+ \beta \lambda = 0.
\end{array}
\]
We consider the following particular solutions.
\begin{itemize}
  \item
    Solution
    \begin{equation}
        \widetilde{U} = 2 \lambda - \frac{Z - \beta \lambda}{Z^\prime}, 
        \qquad
        \widetilde{V} = \widetilde{W} = 0,
        \qquad
        \widetilde{Z} = \widetilde{Z}(\lambda)
    \end{equation}
    leads to
    \begin{equation} \label{A23isol1}
        H = x {\tau}_1 + \frac{\widetilde{Z}(t_1) y^2}{2 t} + y {\tau}_2,
        \qquad
        t_1 = t^2 {\tau}_1,
    \end{equation}
    where ${\tau}_1$ and ${\tau}_2$ are some functions of $t$, and ${\tau}_1$ satisfies
    \[
    t^3 {\tau}_1^{\prime} - 2 t_1 + \frac{\widetilde{Z}(t_1) - \beta t_1}{\widetilde{Z}^{\prime}(t_1)} = 0.
    \]
    In particular, if ${\tau}_1(t) = {\tau}_1^*(t) = t^{-1}$, then one derives
    \begin{equation} \label{A23isol1a}
    H^* = \frac{x}{t}+ \frac{1}{2} (C_1 - \beta \ln{t}) y^2 + y {\tau}_2,
    \end{equation}
    where $C_1$ is a constant.

  \item
  Solution
    \[
        \displaystyle
        \widetilde{U} = \lambda + C_2, 
        \qquad
        \widetilde{V} = 0,
        \qquad
        \widetilde{W} = C_1,
        \qquad
        \widetilde{Z} = \frac{\beta (C_2 - \lambda)}{C_1} + \left((C_1 + 1) \lambda - C_2 \right)^{\frac{1}{C_1+1}},
    \]
    gives
    \begin{equation} \label{A23isol2} 
    \displaystyle
        H = \left(\frac{C_1 y + C_4}{t} + \frac{C_2}{t^2}\right) x
            -\beta \left(\frac{y^3}{6} + \frac{C_4 y^2}{2 C_1}\right)
            + \frac{C_3 ((C_1+1) (C_1 y+C_4) t + C_1 C_2)^{\frac{2 C_1 + 3}{C_1 + 1}}}
                {t^3 C_1^2 (C_1 + 2) (2 C_1 + 3)}
        + y {\tau},
    \end{equation}
    \[
        C_1 \neq -2, -\frac{3}{2}, -1, 0,
    \]
    where $C_1$, ..., $C_4$ are constants, and ${\tau}$ is an arbitrary function of~$t$.
    
    \item
    There is also a solution
    \begin{equation} \label{A23isol3implicit}
        \displaystyle
        \widetilde{U} = 2 \lambda + \left(C_1 \lambda + C_2 - \ln{\theta} \right) \widetilde{V}, 
        \qquad
        \widetilde{V} = \frac{\theta}{\theta^\prime},
        \qquad
        \widetilde{W} = C_1 \widetilde{V},
        \qquad
        \widetilde{Z} = C_1^2 \widetilde{V},
    \end{equation}
    where $C_1$ and $C_2$ are constants, and the function $\theta=\theta(\lambda)$ satisfies the equation
    \begin{equation} \label{A23isol3implicitTheta}
    \frac{(C_2 - \ln{\theta}) \theta^2 \theta^{\prime\prime}}{(\theta^{\prime})^3}
        - \frac{(1 + C_2 - \ln{\theta})\theta}{\theta^{\prime}}
        + \frac{\beta \lambda}{1+C_1^2} = 0.
    \end{equation}
    The solution is written only implicitly through integrals of Bessel functions of the first and second kind and has a quite cumbersome form, so we do not present it here. We will consider this solution in more detail in Section~\ref{sec:numsol}.
\end{itemize}

\subsection{One-dimensional subalgebras}

In the case of one-dimensional subalgebras, system~(\ref{sysRE}) is no longer reduced to a system of ordinary differential equations. In general, this significantly complicates the calculations, and therefore we will consider in detail only some specific solutions here.

\subsubsection{Subalgebra $\{ Y_1 \}$}

\medskip

For the subalgebra~$\{ Y_1 \}$, the functions~$U$, $V$, $W$, and~$Z$ depend on two variables, $y$, and~$h$. 
The reduced system is easily derived by substituting $U_t = V_t = W_t = Z_t = 0$ into~(\ref{sysRE}).

\begin{itemize}
    \item
    First, we consider the specific solution
    \[
    U = V = W = 0,
    \qquad
    Z = \mu(h) - \beta y,
    \]
    where $\mu$ is an arbitrary function of its argument. This leads to the solution
    \begin{equation} \label{A11isol1}
    H = A x - \frac{\beta y^3}{6} + B y^2 + {\tau} y,
    \end{equation}
    where $A$ and $B=\mu(C_1)/2$ are constants, ${\tau}$ is an arbitrary function of~$t$. This is equivalent to the solution~\cite[p.~225, (6.a)]{bk:HandbookLie_v2}.
    
    \item
    Another solution of the reduced resolving system satisfies the equations
    \begin{equation} \label{VarphiOfW}
    V = 0, \qquad
    y W + \varphi(W) - h = 0,
    \end{equation}
    where $\varphi$ is an arbitrary function. If we consider $\varphi(W) = C_1 W$ for some constant $C_1$, there is a solution
    \[
    U = \psi(W) (y + C_1), \qquad
    V = 0, \qquad
    W = \frac{h}{y + C_1}, \qquad
    Z = \Psi\left( 
        \ln{(y+C_1)} + \int \frac{W dW}{\psi(W)}
    \right) - \beta y,
    \]
    where $\psi$ and $\Psi$ are arbitrary functions of their arguments. 
    In particular, if we choose $\psi(z)=z^2$, $\Psi(z) = 2 C_2 z$, then 
    \[
    U = \frac{h^2}{y + C_1}, \qquad Z = 2 C_2 \ln{h} - \beta y,
    \]
    and
    \begin{equation} \label{A11isol2}
    H = C_2 y^2 \ln\left|\frac{y+C_1}{t}\right|
        + C_1 C_2 (2 y+C_1) \ln |y +C_1|
        - \frac{\beta y^3}{6}
        + \frac{x}{t} (y+C_1)
        - \frac{C_2}{2} (3 y + C_1)^2
        + 3 C_2 y^2
        + {\tau} y,
    \end{equation}
    where ${\tau}$ is an arbitrary function of~$t$.
    
    \item
    The last solution, which we consider here in detail, can be found if we assume $V = V_1(y) V_2(h)$ and $W=W(y)$. This gives
    \[
    \def\arraystretch{1.75}
    \begin{array}{c}
    \displaystyle
        U = \frac{\sqrt{h^2 + C_3}}{C_1} \left(
            C_4 e^{C_1 y} + C_5 e^{-C_1 y} + \beta
        \right),
        \qquad
        V = C_1 \sqrt{h^2 + C_3},
        \\
        \displaystyle
        W = 0, 
        \qquad
        Z = \frac{1}{C_1} \left(
            C_4 e^{C_1 y} - C_5 e^{-C_1 y}
        \right),
    \end{array}
    \]
\end{itemize}
where $C_1$, ..., $C_5$ are constants ($C_1 \neq 0$). The corresponding invariant solution is
\begin{equation} \label{A11isol3}
\displaystyle
H = A_1 e^{C_1 (x + {\tau})} + A_2 e^{-C_1 (x + {\tau})}
    + B_1 e^{C_1 y} + B_2 e^{-C_1 y} + \left({\tau}^\prime + \frac{\beta}{C_1^2}\right) y,
\end{equation}
where
\[
    {\tau} = {\tau}(t), \qquad
    2 C_1 A_1 = -1, \qquad
    2 C_1 A_2 = C_3, \qquad
    C_1^3 B_1 = C_4, \qquad
    C_1^3 B_2 = -C_5.
\]
This solution generalizes the known solution~\cite[p.~225, (6.c)]{bk:HandbookLie_v2}.

\subsubsection{Subalgebra $\{ Y_2 \}$}

\medskip

In this case the functions $U$, $V$, $W$, and $Z$ depend on $t$ and $h$. We consider the following solution
\[
\def\arraystretch{2}
\begin{array}{c}
U = q(t) - (\ln W)_t \, h,
\qquad
V = 0,
\qquad
W = W(t),
\\
\displaystyle
Z = 
    Q\left( h I + \int q I \, dt \right)
    -\left\{
        \left( h I + \int q I \, dt \right) \int \frac{\, dt}{I}
        - \int \frac{1}{I} \left(\int \frac{q \, dt}{I} \right) \, dt
\right\} \beta,
\end{array}
\]
where $q$ and $Q$ are arbitrary functions of their arguments, and
\[
I = W \exp \left(-\int W dt \right).
\]
In general it does not seem that the latter can be integrated, and therefore we consider only the specific case
\[
W = W^* = \frac{1}{1 + C_1 e^{-t}} = \sigma_1(t), 
\qquad
q = C_2,
\qquad
Q = 0,
\]
which leads to the particular solution
\begin{equation} \label{A12isol1}
H^* = \left(
    \left(y + C_2 (C_1 e^{-t} - t) + C_3\right) x
    - \frac{\beta y^3}{6} - \frac{\beta}{2}  (C_2 (C_1 t e^{-t}-1)+C_3) y^2
\right) \sigma_1(t) + \tau y,
\end{equation}
where $C_1$, $C_2$, and $C_3$ are constants, and~$\tau$ is a function of~$t$.
Notice that this solution is very similar to~(\ref{A22isol2}).

\subsubsection{Subalgebra $\{ Y_3 \}$}

\medskip

In this case,
\[
U = t^{-3} \widetilde{U}(z , w), \quad
V = t^{-1} \widetilde{V}(z , w), \quad
W = t^{-1} \widetilde{W}(z , w), \quad
Z = t^{-1} \widetilde{Z}(z , w), \quad
z = t y, \quad
w = t^2 h.
\]

Assumption $\widetilde{V} = \textrm{const}$ brings to polynomial solutions which are of not much interest.

If we assume $\widetilde{V} = \widetilde{V}(w)$, $\widetilde{W} = i \widetilde{V}(w)$, $i = \sqrt{-1}$, then the remaining functions are
\[
\displaystyle
\widetilde{Z} = -\beta z - \widetilde{V},
\qquad
\widetilde{U} = \left(k-\frac{\beta z^2}{2}\right) \widetilde{V} 
    + \widetilde{V} \int  \left( {\widetilde{V}}^{-1} - 2 w \widetilde{V}^{-2} \widetilde{V}^{\prime}\right) dw,
\]
where $k$ is some complex-valued constant.

For example, we consider the simplest case $\widetilde{V}(w)=\widetilde{V}^*=w$. 
Then, the following solution can be derived
\begin{equation} \label{A13isol1}
\displaystyle
H^* = t^{-3} A^{-t} e^{t (x + i y + \tau)}
    - \frac{\beta y^3}{6}
    + \tau^\prime y,
\end{equation}
where $A$ is a constant and $\tau$ is a function of $t$.

\subsubsection{Subalgebra $\{ Y_1 + \alpha Y_2 \}, \; \alpha = \pm 1$}

\medskip

Here the functions $U$, $V$, $W$, and $Z$ depend on $h$ and $\lambda = y - \alpha t$. 
Consideration of this case is quite similar to the case of subalgebra~$\{ Y_1 \}$, 
thus we just provide two particular invariant solutions.
\begin{equation} \label{A14isol1}
\displaystyle
H_1 = A_1 e^{A_3 (x + \tau_1)} + A_2 e^{-A_3 (x+\tau_1)} + \left(\tau_1^{\prime}+ \frac{\beta}{A_3^2}\right) \lambda,
\end{equation}
and
\begin{equation} \label{A14isol2}
H_2 = C_1 x - \frac{C_1 \beta}{6 (C_1 - \alpha)} \lambda^3 + C_2 \lambda^2 + \tau_2 \lambda,
\end{equation}
where  $A_j$, $C_k$ are constants, $\tau_1$, $\tau_2$ are arbitrary functions of~$t$.
Notice that, in contrast to (\ref{A11isol3}), there is no~$e^\lambda$-terms in this case.

\medskip

Another case corresponds to the solution~(cf.~(\ref{VarphiOfW}))
\[
V = 0, \qquad
\lambda W + \varphi(W) - h = 0,
\]
The choice of a particular function $\varphi$ determines a set of invariant solutions. If we assume $\varphi(f) = C_1 f$, then
\[
U = (\lambda + C_1) \psi(W) + \alpha W.
\]
Repeating the arguments carried out for the subalgebra~$\{ Y_1 \}$, we finally arrive at the solution
\begin{equation} \label{A14isol3}
H_3 =  \left(
        \frac{x}{t} 
        +\tau 
    \right) \lambda
    + \left(\beta (C_1 - \alpha t)- \alpha C_2 \ln{t}- \frac{C_1 C_2}{t}\right) \frac{\lambda^2}{2}
    - \left(\beta + \frac{C_2}{t}\right) \frac{\lambda^3}{6}
    + C_1 \frac{x}{t},
\end{equation}
where $C_2$ is a constant and $\tau$ is an arbitrary function of~$t$.

\subsection{Some numerical issues}
\label{sec:numsol}

Here we consider solution~(\ref{A23isol3implicit}), where equation~(\ref{A23isol3implicitTheta}) can be solved only numerically. 
The peculiarity of such solutions is that, having obtained a numerical solution for the function $\theta$, we still have need to integrate the automorphic system. Thus, we have an automorphic system, on the left side of which there are differential relations, while the right side can only be obtained numerically. A similar situation does not arise when directly reducing the original equation on a subgroup and this issue has to be handled in some way. In this specific case, we can significantly simplify the problem by setting the constant $C_1=0$. (We also set $C_2=0$, as this parameter has little effect on the calculation results.) Then, we have
\[
H_{tx}  = \frac{1}{t^3} \, \left(\frac{\theta \ln {\theta}}{\theta^\prime} - 2 \lambda\right), 
        \qquad
H_{xx} = \frac{1}{t} \frac{\theta}{\theta^\prime},
        \qquad
H_{xy} = 0,
\qquad
H_{yy} =0.
\]
For brevity, we assume $H_y = 0$. Otherwise, $H$ is a linear function of~$y$, which is not of much interest. 

\begin{figure}[H]
    \centering
    $\vcenter{\hbox{\includegraphics[width=0.42\linewidth]{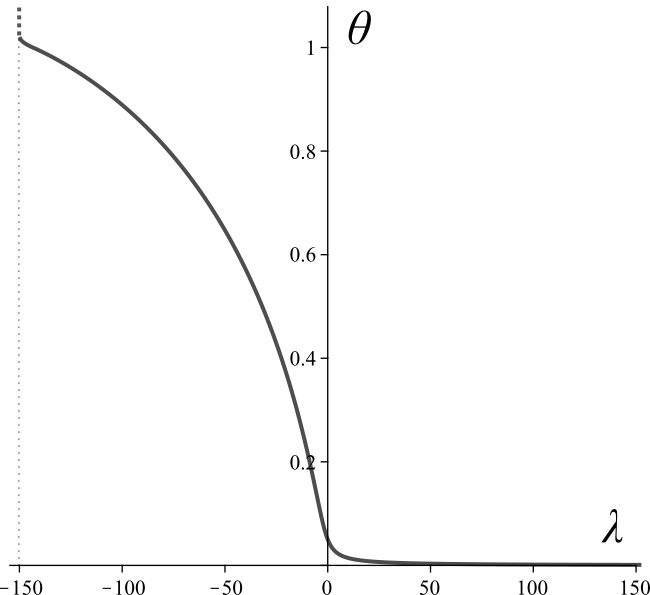}}}$
    \hfill
    $\vcenter{\hbox{\includegraphics[width=0.48\linewidth]{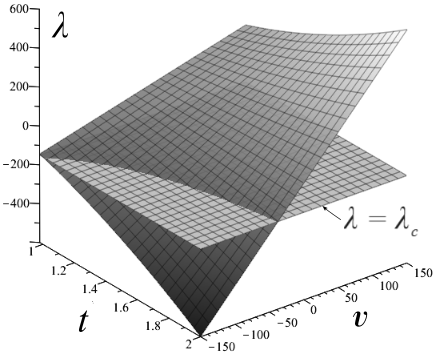}}}$
    \caption{Left: Numerical solution for $\theta(\lambda)$. 
    Right: The plot for~$\lambda=t^2 v$ near the region $\lambda<\lambda_c$, where the function~$\theta(\lambda)$ has a singularity.
    }
    \label{fig:fig1}
\end{figure}

Recall that $H_x$ is the fluid velocity component~$v$ in the direction of the~$y$ axis. Then, from the latter equations we get
\[
v_t  = \frac{1}{t^3} \, (\widetilde{V} \ln \theta - 2 \lambda), 
\qquad
v_x = \frac{\widetilde{V}}{t},
\qquad
\widetilde{V} = \frac{\theta}{\theta^\prime}.
\]
Now we can at least investigate the behavior of the~$y$-component of the velocity.
For this purpose, we consider the simplest explicit finite-difference scheme
\begin{equation}
\label{partNumSol}
\displaystyle
v_{m}^{n+1} = v_m^n 
    + \frac{\tau}{ \sigma} \frac{\left(v_{m+1}^n - v_m^n\right) \ln{\theta^n_m}}{t_n^2}
    - \frac{2 \tau \lambda^n_m}{t_n^3},
\qquad
v_{m+1}^n = v_m^n + \frac{\sigma}{t_n} \frac{\theta^n_m}{(\theta^\prime)^n_m},
\qquad
\lambda^n_m = t_n^2 v^n_m,
\end{equation}
where $m$ and $n$ are indexes of the orthogonal grid along $x$ and $t$ axes, $\sigma = x_{i+1} - x_i$ and $\tau = t_{i+1} - t_i$ are 
constant steps.\footnote{As the scheme is explicit, it is reasonable to get~$0<\tau < \sigma^2$. We choose $\sigma = 0.02$ and $\tau = 0.0625 \sigma^2$.} Here we also denoted
\[
\theta^n_m = \theta(\lambda^n_m),
\qquad
(\theta^\prime)^n_m = \theta^\prime(\lambda^n_m).
\]
First, we set the Cauchy problem $\theta(\lambda_0) = \theta_0$, $\theta^\prime(\lambda_0) = \theta_1$ for some constants $\lambda_0 = t_0^2 v_0$, $\theta_0$ and $\theta_1$. Figure~\ref{fig:fig1}~(left) gives a numerical solution for the problem with the parameters~$t_0 = 0.1$, $v_0 = 1$, $\theta_0=0.05$, $\theta_1=-0.01$, $\beta=4$. This solution allows us to specify an automorphic system. 
Notice that at $\lambda = \lambda_c \approx -150$, the solution has a singularity, which is not appropriate for solving~(\ref{partNumSol}). 

\begin{figure}[ht]
    \centering
    $\vcenter{\hbox{\includegraphics[width=0.48\linewidth]{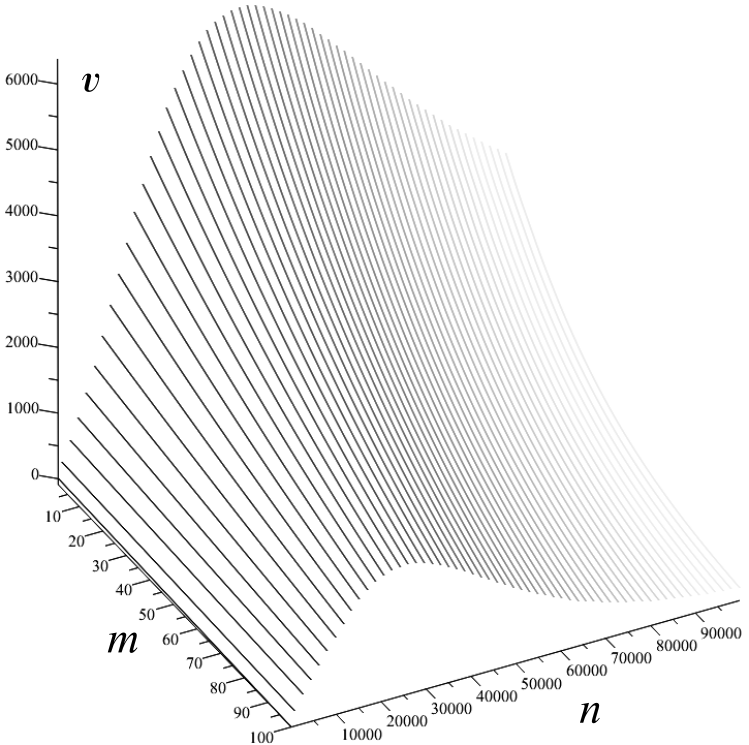}}}$
    \hfill
    $\vcenter{\hbox{\includegraphics[width=0.48\linewidth]{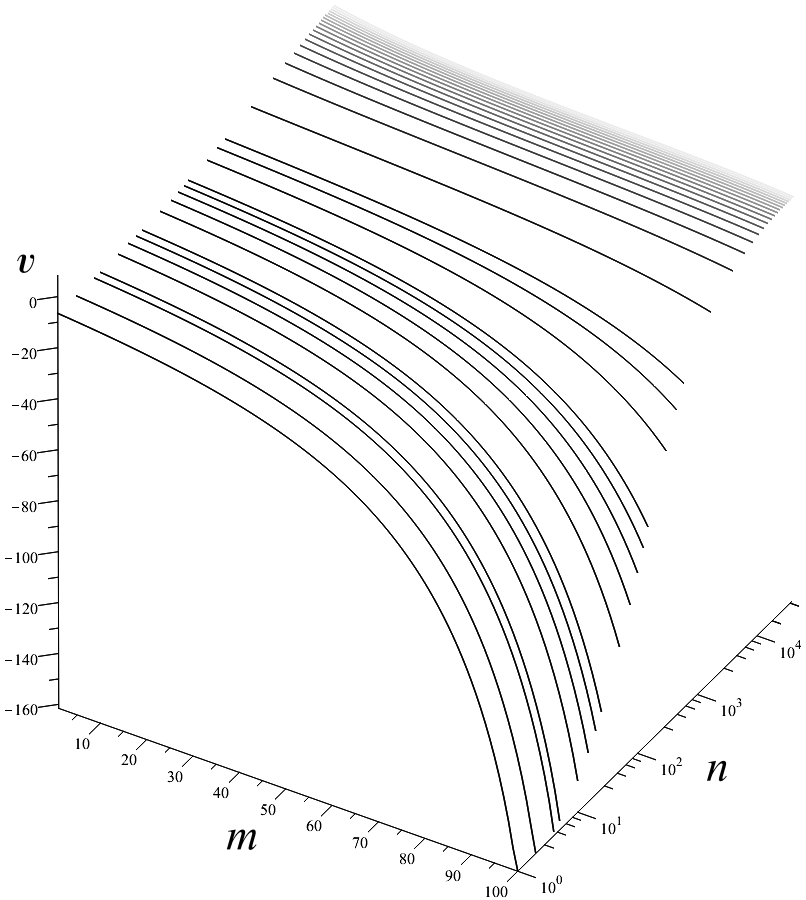}}}$
    \caption{Numerical solutions for $v^n_m=v(m \sigma, t_0 + n \tau)$. 
    Left: $\lambda > 0$. 
    Right: $\lambda \approx \lambda_c$.}
    \label{fig:fig2}
\end{figure}

Considering $\lambda_0 > 0$ (starting with $v^0_0 = 1$, $t_0=0.1$), we obtain the solution~$v^n_m$ presented in Figure~\ref{fig:fig2}~(left). 
For the given parameters, it can be seen that~$v$ decreases almost linearly with increasing~$x$ for any time~$t$.
On the contrary, in the direction of the time axis the behavior of~$v$ is essentially nonlinear.

Figure~\ref{fig:fig2}~(right) demonstrates the behavior of the velocity component~$v$ near the singularity. 
To derive this picture, we start counting at the point $v^0_0=-6.55$, $t_0=0.98$. In this case, $\lambda$ at the boundary of the computational domain approaches~$\lambda_c$. Here one sees that over time the velocity value `levels off', approaching zero, and then the picture becomes similar to~Figure~\ref{fig:fig2}~(left).

A similar strategy can be applied to other numerical solutions of the resolving system, although the way to obtain the velocity component $u=-H_y$ in many  cases can be much more problematic.

\subsection{Discussion}
\label{sec:disc}

Thus, by means of the group foliation approach, we have derived some particular invariant solutions to equation~(\ref{EqMain}). 
In one of the cases, we are able to integrate the resolving system only numerically. 
As it was established in~\cite{bk:Golovin_strat[2003]}, invariant solutions obtained using group foliations can be of a more general form than invariant solutions obtained by the standard method. Invariant solutions of equation~(\ref{EqMain}) have been studied in the papers~\cite{bk:Katkov_Geostr[1965]}, \cite{bk:Katkov_Geostr[1966]}~(they are also listed in~\cite{bk:HandbookLie_v2}, which is more convenient to cite), \cite{HUANG2004428,Bihlo_2009a,Bihlo_2009b}. 
Invariant solutions for a system of a more general form have been also obtained in~\cite{bk:BihloPop_Geostr_InvSols[2011]}. Based on the solutions given in~\cite{bk:Katkov_Geostr[1965],bk:Katkov_Geostr[1966],Bihlo_2009a,Bihlo_2009b,bk:BihloPop_Geostr_InvSols[2011],bk:HandbookLie_v2} , one can conclude that the classes of solutions obtained there and the solutions derived in the present publication overlap to a large extent. 
Mostly, these solutions reduce to polynomials (up to the degree of three), exponential, harmonic (including Rossby waves solutions~\cite{HUANG2004428}), and logarithmic functions and their linear combinations. 
Apparently, solution~(\ref{A23isol2}) cannot be reduced to any of the solutions given in~\cite{bk:Katkov_Geostr[1965],bk:Katkov_Geostr[1966],Bihlo_2009a,Bihlo_2009b,bk:BihloPop_Geostr_InvSols[2011],bk:HandbookLie_v2} and it seems to be new.

Comparing the solutions we obtained here with the invariant solutions given in~\cite{bk:HandbookLie_v2} for $\beta=1$, one can see that the solutions derived here through group foliation approach are often more general. For example, the solutions called ``configurations of type of narrow gullies and crests''~\cite{bk:HandbookLie_v2} are obtained here in a more general form:~for the subalgebra~$\{ Y_1 \}$ we found a solution of type~\cite[p.~225, (6. a)]{bk:HandbookLie_v2} and \cite[p.~225, (6.c)]{bk:HandbookLie_v2}, and for the subalgebra $\{ Y_1 - Y_2 \}$ there is a solution of type~\cite[p .~225, (6.b)]{bk:HandbookLie_v2}. A solution of the form~\cite[p.~224, (4)]{bk:HandbookLie_v2} was obtained for the subalgebra~$\{ Y_1, Y_2 \}$. (See the sections on the corresponding subalgebras for more details.)

Notice that additional solutions can be derived if one imposes some restrictions on the variable $h=H_x$. 
Some of the listed solutions are obtained only under the condition~$h=\textrm{const}$, and, for example, solution~\cite[p.~223, (1)]{bk:HandbookLie_v2} is obtained for the subalgebra~$\{ Y_1, Y_2 \}$ only under the assumption~$H = \phi(t,x)$ (i.e., $H_y=0$). 
Thus, if derivatives of the function sought are chosen as new dependent invariant variables for a group foliation, then in order to obtain an extended set of exact solutions, special cases should be considered when these derivatives become constants, depend on an incomplete set of independent variables, etc.

In addition, as we saw in the previous section, if the solution to a resolving system can only be obtained numerically, the procedure for solving an automorphic system can become significantly more complicated (for example, it may require the construction of a finite-difference scheme).

To conclude this discussion, we list the invariant solutions obtained here in Table~\ref{tab:invsols} to make them easier to find in the text.
\bgroup
\def\arraystretch{1.5}
\begin{longtable}[c]{|c|l|}
\hline
Subalgebra of~(\protect\ref{algY}) & Particular invariant solutions \\ 
\hline
\hline
\endfirsthead
\caption*{\textit{(Continuation of Table~\protect\ref{tab:invsols})}}\\
\hline
Subalgebra & Particular invariant solutions \\ 
\hline
\hline
\endhead
$\{ Y_1, Y_2, Y_3 \}$ &  Eq. (\protect\ref{H_A3}) \\ 
$\{ Y_1, Y_2 \}$ &  Eqs. (\protect\ref{GaurvitzSol}), (\protect\ref{A21isol1}), (\protect\ref{A21isol2}) \\ 
$\{ Y_1, Y_3 \}$ &  Eqs. (\protect\ref{A22isol1}),  (\protect\ref{A22isol2}) \\  
$\{ Y_2, Y_3 \}$ &  Eqs. (\protect\ref{A23isol1}), (\protect\ref{A23isol1a}), (\protect\ref{A23isol2}), (\protect\ref{A23isol3implicit}) \\ 
$\{ Y_1 \}$ &  Eqs. (\protect\ref{A11isol1}), (\protect\ref{A11isol2}), (\protect\ref{A11isol3}) \\ 
$\{ Y_2 \}$ &  Eq. (\protect\ref{A12isol1}) \\ 
$\{ Y_3 \}$ &  Eq. (\protect\ref{A13isol1}) \\ 
$\{ Y_1 + \alpha Y_2 \}$, $\alpha = \pm 1$  &  Eqs. (\protect\ref{A14isol1}), (\protect\ref{A14isol2}), (\protect\ref{A14isol3})  \\ \hline
\caption{Some particular invariant solutions $H(t,x,y)$ found, up to transformations~(\protect\ref{XinfGrTransform}).}
\label{tab:invsols}\\
\end{longtable}
\egroup

\section{Conservation laws of equation~(\ref{EqMain})}
\label{sec:sec4}



\bigskip

\textbf{Disclaimer.} \emph{A considerable time after the publication of previous versions of this preprint, the author became aware that the results presented in this section were obtained earlier in~\cite{bk:Jamal2020} (it is also worth mentioning some results from~\cite{DGAndrews1983}). However, in order to preserve the integrity of the presentation, the author provides here a text devoted to conservation laws. The author does not claim originality of the results presented in this section.}

\bigskip

Although it is known that equation~(\ref{EqMain}) can be derived from the system of incompressible two-dimensional Euler equations in a rotating reference frame in the absence of external forces~\cite{bk:BihloPop_Geostr_numeric[2019]}, this does not mean that the set of conservation laws of equation~(\ref{EqMain}) is exhausted by the conservation laws of Euler equations. The introduction of the velocity potential~$(-H_y, H_x)$ may also lead to some new conservation laws in potential coordinates that cannot be directly expressed in Eulerian coordinates. Similar examples are often observed for equations of hydrodynamic type in Lagrangian coordinates (see, e.g,~\cite{bk:KaptsovMeleshko2019,bk:GilmanModelKapMelDor[2022]}). 


\smallskip

Recall that the local conservation law of equation~(\ref{EqMain}) is the following divergent expression that vanishes on solutions of~(\ref{EqMain})
\[
D_t Q^t + D_x Q^x + D_y Q^y = 0 ,
\]
where $Q^t$, $Q^x$, $Q^y$ are its conserved quantities. 
By the \emph{order} of a conservation law we will mean the highest order of derivatives on which its conserved quantities depend.

Any conservation law of~(\ref{EqMain}) can also be written in the form
\begin{equation} \label{CLLambdaF}
D_t Q^t + D_x Q^x + D_y Q^y \equiv \Lambda F.
\end{equation}
where~$F$ is the left hand side of~(\ref{EqMain}) and $\Lambda$ is called a conservation law multiplier. 
If a conservation law multiplier is known, then finding the corresponding conserved quantities becomes a purely technical problem.

\smallskip

In contrast to partial differential equations of even orders, for odd orders, the situation usually becomes much more complicated, since the nature of correspondence between symmetries and conservation law multipliers is much less understood. There are various approaches here, such as the use of adjoint equations~\cite{DorKapKozWin2015} and adjoint~Lagrangian~\cite{bk:Ibragimov2011}, the action of symmetries on the components of already known conservation laws~\cite{bk:Anco_AntiIbragimov[2017]}, the direct method~\cite{bk:AncoBluman1997,bk:BlumanCheviakovAnco}, etc. 
A detailed discussion of the issues of finding conservation laws and their multipliers, especially utilizing the direct method, can be found in~\cite{bk:BlumanCheviakovAnco}. There is also an overview of all the main currently known approaches in~\cite{bk:NazMahomedMason2008}.

\subsection{Conservation laws of order two}

To begin with, we consider conservation laws with multipliers depending on derivatives of no higher than the second order. 
This seems reasonable since multipliers that are usually considered are of order one less than the order of the equation. 
In addition, this assumption leads to a feasible amount of calculations.

To find such conservation laws, we use a straightforward approach, which consists of two steps. At the first step we find all possible conservation law multipliers of order two or lower by utilizing the direct method~\cite{bk:AncoBluman1997}. The second step is to determine the conservation laws corresponding to the obtained multipliers by direct calculations.

\medskip

Recall that the main idea of the direct method is to apply the~Euler operator 
\begin{equation} \label{OpEuler}
\displaystyle
\mathcal{E} = \frac{\partial}{\partial H}
 - D_t\left(\frac{\partial}{\partial H_t} \right)
 - D_x\left(\frac{\partial}{\partial H_x} \right)
 - D_y\left(\frac{\partial}{\partial H_y} \right)
 + D_t^2 \left(\frac{\partial^2}{\partial H_{tt}} \right)
 + D_t D_x \left(\frac{\partial^2}{\partial H_{tx}} \right)
 + \dots
\end{equation}
to the expression $\Lambda F$. 
It is known that any divergent expression identically vanishes under the action of~(\ref{OpEuler}), thus we consider the identity
\begin{equation} \label{EulerOpEqn}
\mathcal{E} (\Lambda F) \equiv 0.
\end{equation}
If $\Lambda F$ is a conservation law (see (\ref{CLLambdaF})), then the corresponding multiplier~$\Lambda$ can be found in this way.

\medskip

Now we consider conservation law multipliers of order two, namely
\begin{equation} \label{Lambda2}
\Lambda = \Lambda(t,x,y,H,H_t, H_x, H_y, H_{tt}, H_{tx}, H_{ty}, H_{xx}, H_{xy}, H_{yy}).
\end{equation}
Applying~(\ref{OpEuler}) and splitting the resulting equation~(\ref{EulerOpEqn}) with respect to the derivatives of orders three and four, 
we derive an overdetermined system of partial differential equations on $\Lambda$. 
Deriving a solution to this system is not particularly difficult, but it requires significant amount of computations, 
so here we present only the final result. The conservation law multiplier~(\ref{Lambda2}) of (\ref{EqMain}) has the form
\begin{equation} \label{LambdaGen}
\Lambda = \frac{x^2 + y^2}{2} \,{\tau}_1 + x {\tau}_2 + y {\tau}_3 + {\tau}_4 + c_1 H + c_2 (H_{xx} + H_{yy}) + \Phi(H_{xx} + H_{yy} + \beta y),
\end{equation}
where $c_1$ and $c_2$ are constants, $\Phi$ is an arbitrary smooth function of its argument, and ${\tau}_1$, ..., ${\tau}_4$ are arbitrary smooth functions of~$t$ satisfying the relations
\begin{equation} \label{clConds}
\beta {\tau}_1 = 0,
\qquad
\beta {\tau}_2 + 2 {\tau}_1^\prime = 0.
\end{equation}
Thus, in case $\beta \neq 0$ one derives
\[
\Lambda|_{\beta \neq 0} = y {\tau}_3 + {\tau}_4 + c_1 H + c_2 (H_{xx} + H_{yy}) + \Phi(H_{xx} + H_{yy} + \beta y).
\]
In case $\beta = 0$, the multiplier becomes
\[
\Lambda^0 = \Lambda|_{\beta = 0} = \frac{x^2 + y^2}{2} \, c_3 + x {\tau}_2 + y {\tau}_3 + {\tau}_4 + c_1 H + \Phi(H_{xx} + H_{yy}),
\]
where $c_3$ is constant.

\medskip

Then, the second step is to find the conserved quantities $Q^t$, $Q^x$, and $Q^y$ satisfying the identity~(\ref{CLLambdaF}) 
assuming~(\ref{LambdaGen}) and~(\ref{clConds}). 
In this case, the calculations also require considerable effort, but can be significantly simplified if one looks for the quantities of a polynomial form.
By this way, for an arbitrary $\beta$ one arrives at the following list of conservation laws.
\begin{multline} \label{CLmomentum}
\Lambda_1 = {\tau_4}(t): \qquad
J_1 = D_t\left(\tau_4 H_{yy} \right) 
- D_x\left( \left( (H_{xx}+H_{yy}) H_{y} - H_{tx} - \beta H \right) \tau_4 \right) \\
+ D_y\left( \left((H_{xx}+H_{yy}) H_{x}\right) \tau_4 \right) = 0.
\end{multline}
\begin{multline} \label{CLnoname1}
\Lambda_2 = y {\tau_3}(t): \qquad
J_2 = D_t\left( y {\tau_3} H_{yy} \right) 
+ D_x\left( 
    \left(
        H_{y}^2 + (H_{x} H_{xy} - H_{y} H_{xx} + H_{tx} + \beta H) y
    \right) {\tau_3} 
\right)  \\
+ D_y\left( 
    (H-y H_{y}) {\tau_3}^{\prime} 
    +\left((H_{x} H_{yy}- H_{y} H_{xy} ) y - H_{x} H_{y} \right) {\tau_3} 
\right) = 0,
\end{multline}
\begin{multline} \label{CLenergy}
\Lambda_3 = H: \qquad
J_3 = D_t\left( {\frac{H_{x}^2 + H_{y}^2}{2}}\right) 
+ D_x\left(((H_{xx}+H_{yy}) H_{y} -H_{tx}) H -{\frac{\beta H^2}{2}}\right)  \\
- D_y\left( (H_{ty} + (H_{xx}+H_{yy}) H_{x}) H \right) = 0,
\end{multline}
\begin{multline} \label{CLabsVel}
\Lambda_4 = \Phi(H_{xx}+H_{yy}+\beta y): \quad
J_4 = D_t\left( \Phi(H_{xx}+H_{yy}+\beta y) \right)  
        - D_x\left( \Phi(H_{xx}+H_{yy}+\beta y) H_{y} \right) \\
        + D_y\left( \Phi(H_{xx}+H_{yy}+\beta y) H_{x} \right) = 0,
\end{multline}
\begin{multline} \label{CLvorticity}
\Lambda_{4a} = H_{xx} + H_{yy}: \qquad
J_{4a} = D_t\left( 
        {\frac{(H_{xx} + H_{yy})^2}{2}}
        +\beta t (H_{yy} + H_{xx}) H_{x} 
    \right) \\
- D_x\left( 
    \beta t (H_{t} H_{yy}+H_{x} H_{tx})+{\frac{(H_{xx}+H_{yy})^2 H_{y} }{2}}
    \right)  \\ 
+ D_y\left( 
    \frac{(H_{xx}+H_{yy})^2 H_{x}}{2}
    -\beta t (H_{x} H_{ty} - H_{t} H_{xy}) 
\right) = 0.
\end{multline}
\medskip

In case $\beta=0$, there are two additional conservation laws:
\begin{multline} \label{CLb0_1}
\Lambda_5^0 = x {\tau_2}(t): \qquad
J_5^0 = D_x\left( 
    \left((H_{x} H_{xy}-H_{y} H_{xx}+H_{tx}) x - H H_{xy}-H_{t} \right) {\tau_2} 
\right)  \\
+ D_y\left( 
    \left((H_{x} H_{yy}-H_{y} H_{xy}+H_{ty}) x + H H_{xx} \right) {\tau_2} 
\right) = 0,
\end{multline}
\begin{multline} \label{CLb0_2}
\Lambda_6^0 = \frac{x^2+y^2}{2}: \qquad
J_6^0 = D_t\left( 
    x H_{x}
    + \left(
        (H_{y} H_{xy} - H_{x} H_{yy}) y
        + H H_{xy}
    \right) t
     + 3 H 
\right) \\
+ D_x\left( 
    {\frac{x^2 + y^2}{2}} \left(
        H_{tx} + H_{x} H_{xy} - H_{y} H_{xx}
    \right)
    -2 y (t H_{y}-x) H_{ty}
    -x H H_{xy}
    -t (H H_{ty}+H_{t} H_{y}) 
\right) \\
+ D_y\left( 
    {\frac{x^2 + y^2}{2}} \left( H_{ty}  + H_{x} H_{yy} - H_{y} H_{xy}\right)
    +\left( (H_{x} H_{ty} + H_{y} H_{tx}) t - 2 x H_{tx} -3 H_{t}\right) y
    + x H H_{xx}
\right) = 0.
\end{multline}


Some of the listed conservation laws have clear physical interpretations.
For~${\tau_4}(t)=\textrm{const}$, the conservation law (\ref{CLmomentum}) is just the original equation~(\ref{EqMain}) which corresponds to the conservation law of momentum in Eulerian coordinates. The conservation laws of energy and absolute velocity are given by~(\ref{CLenergy}) and (\ref{CLabsVel}), respectively. 
One might also associate~(\ref{CLvorticity}) with the vorticity preservation.

\begin{remark}
The conservation law $J_{4a}$ is not independent:
\[
\displaystyle
J_{4a} = J_2|_{\tau_3(t) = -\beta} + \frac{1}{2} J_4|_{\Phi(z) = z^2} + D_0,
\]
where $D_0$ is a trivial conservation law (a divergence expression which is identically zero). 
Nevertheless, we have presented this conservation law here because it might be interesting in its own right. 
\end{remark}

\begin{remark}
The conservation law~$J_1$ can be reduced to the following equivalent conservation law in `stationary' form
\[
\widetilde{J}_1 = 
D_x\left( \left( H_{tx} + H_{x} H_{xy} - H_{y} H_{xx} + \beta H \right) \tau_4 \right) 
+ D_y\left( \left(H_{x} H_{yy} - H_{y} H_{xy} + H_{ty}\right) \tau_4 \right) = 0.
\]
The corresponding conservation law multiplier remains the same.
\end{remark}

\subsection{Conservation laws of higher orders}

Knowing the symmetries of equation~(\ref{EqMain}), based on the conservation laws obtained in the previous section, it is possible to obtain a number of higher order conservation laws. This can be done by a fairly straightforward method, described, e.g., in~\cite{bk:Anco_AntiIbragimov[2017]}.\footnote{The method proposed in~\cite{bk:Ibragimov2011} should lead to similar results, but in a much more complex manner.} 

Recall that any generator
\[
\displaystyle
    X = \xi^t \partial_t 
        + \xi^x \partial_x
        + \xi^y \partial_y
        + \eta \partial_H
        + \dots
\]
can be represented in characteristic (or evolutionary~\cite{bk:Olver[1986]}) form
\[
\displaystyle
\hat{X} = \hat{\eta} \partial_H 
    + D_t(\hat{\eta}) \partial_{H_t}
    + D_x(\hat{\eta}) \partial_{H_x}
    + D_y(\hat{\eta}) \partial_{H_y}
    + D_t^2(\hat{\eta}) \partial_{H_{tt}}
    + D_t D_x(\hat{\eta}) \partial_{H_{tx}}
    + \dots,
\]
where
\[
\hat{\eta} = \eta - \xi^t H_t  - \xi^x H_x - \xi^y H_y
\]
is called its characteristic.
Suppose 
\[
D_t Q^t + D_x Q^x + D_y Q^y = 0
\]
is a known conservation law of~(\ref{EqMain}). 
Then, according to~\cite{bk:Anco_AntiIbragimov[2017]}, for any generator $X$ admitted by~(\ref{EqMain})  
\[
D_t(\hat{X}(Q^t)) + D_x(\hat{X}(Q^x)) + D_y(\hat{X}(Q^y)) = 0
\]
is again a conservation law.

Thus, using the generators of algebras~(\ref{alg}) and~(\ref{alg0}), and the conservation laws from the previous section, new conservation laws of equation~(\ref{EqMain}) can be generated. 
A significant disadvantage of the described approach is that such conservation laws usually have an order equal to or higher than the order of the original equation and for partial differential equations they are reduced to conservation laws of lower orders only in rare cases. 
As a result, it is often not possible to write such conservation laws in the form~(\ref{CLLambdaF}). 
Instead, they are written as
\begin{equation} \label{CLExtForm}
D_t Q^t + D_x Q^x + D_y Q^y \equiv \Lambda^{0,i} \Psi_i(F) + \Lambda^j \mathcal{D}_j {F},
\end{equation}
where $\Lambda^{0,i}$, $\Lambda^j$ are multipliers, $\Psi_i$ are functions of $F$, and $\mathcal{D}_j$ are some differential operators.
Moreover, the physical meaning and practical applicability of such conservation laws usually remains unclear. 

To set an example, we present here the conservation laws, which are derived from the conservation 
law of energy~(\ref{CLenergy}) using the generators of algebra~(\ref{alg}).

The generators $\hat{X}_1$ and $\hat{X}_2$ give the conservation laws equivalent to $D_t(H F)$ and $D_y(H F)$ which are just total differentiations of the conservation law~(\ref{CLenergy}). The conservation laws given by the generators $\hat{X}_3$ and $\hat{X}_\infty$ are of more interest and have the following form.
\begin{multline*}
D_t\left\{(t  H_{tx}-x  H_{xx}-y  H_{xy}+2 H_{x}) H_{x}+(t  H_{ty}-x  H_{xy}-y  H_{yy}+2 H_{y}) H_{y}\right\}
\\
+ D_x\left\{
    (t H_{t}-x H_{x}-y H_{y}+3 H) ({\zeta} H_{y}-\beta H- H_{tx})
    +(t  H_{ty}-x  H_{xy}-y  H_{yy}+2 H_{y}) H {\zeta}
    \right.
    \\
    \left.
    +(t H_{txx}-x H_{xxx}-y H_{xxy}+ H_{xx}) H H_{y}
    +(t H_{tyy}-x H_{xyy}-y H_{yyy}+ H_{yy}) H H_{y}
    \right.
    \\
    \left.
    -(t H_{ttx}-x H_{txx}-y H_{txy}+3  H_{tx}) H
\right\}
- D_y\left\{
    (t H_{t}-x H_{x}-y H_{y}+3 H) ({\zeta} H_{x}+ H_{ty})
    \right.
    \\
    \left.
    + (t  H_{tx}-x  H_{xx}-y  H_{xy}+2 H_{x}) H {\zeta}
    + (t H_{txx}-x H_{xxx}-y H_{xxy}+ H_{xx}) H H_{x}
    \right.
    \\
    \left.
    + (t H_{tyy}-x H_{xyy}-y H_{yyy}+ H_{yy}) H H_{x}
    + (t H_{tty}-x H_{txy}-y H_{tyy}+3 H_{ty}) H
\right\}
\\
= (x H_x + y H_y - t H_t - 5 H) F + H (x D_x F + y D_y F - t D_t F) = 0,
\end{multline*}
\begin{multline*}
D_t\left\{
    -f  H_{xx} H_{x} - ({f^\prime}+f  H_{xy}) H_{y}
\right\}
+ D_x\left\{
    (g-y {f^\prime}-f H_{x}) ({\zeta} H_{y} - \beta H- H_{tx})
    - ({f^\prime}+f  H_{xy}) H\zeta 
    \right.
    \\
    \left.
    -f H  H_{y} \zeta_x 
    +({f^\prime}  H_{xx}+f H_{txx}) H
\right\}
+D_y\left\{
    f H \left(\zeta H_{x}\right)_x 
    - (g-y {f^\prime}-f H_{x}) ({\zeta} H_{x} + H_{ty})
    \right.
    \\
    \left.
    + ({f^{\prime\prime}}+{f^\prime}  H_{xy}+f H_{txy}) H
\right\}
= (y f^\prime + f H - g) F + f H D_x F = 0,
\end{multline*}
where $\zeta = H_{xx} + H_{yy}$.

\medskip

Since the conservation laws obtained in this way are rather cumbersome, for the remaining conservation laws we only present their representation in terms of their multipliers. Conservation laws that are not just differentiations of conservation laws of lower orders are given in Table~\ref{tab:cls}.

\bgroup
\def\arraystretch{1.5}
\begin{longtable}[c]{|lcc|}
\hline
\multicolumn{1}{|l|}{`Basis' conservation law}  & \multicolumn{1}{l|}{Symmetry}   & Derived conservation law\\ 
\hline
\hline
\endfirsthead
\caption*{\textit{(Continuation of Table~\protect\ref{tab:cls})}}\\
\hline
\multicolumn{1}{|l|}{`Basis' conservation law}  & \multicolumn{1}{l|}{Symmetry}   & Derived conservation law\\ 
\hline
\hline
\endhead
\multicolumn{3}{|c|}{arbitrary $\beta$}        \\ \hline
\multicolumn{1}{|l|}{$J_1$, Eq.(\protect\ref{CLmomentum})} & \multicolumn{1}{c|}{$\hat{X}_3$} &  $2 \tau_4^\prime F + \tau_4^\prime (t D_t F - x D_x F - y D_y F)$ \\  \hline
\multicolumn{1}{|l|}{$J_2$, Eq.(\protect\ref{CLnoname1})} & \multicolumn{1}{c|}{$\hat{X}_3$} &  $2 y \tau_3^\prime F + y \tau_3^\prime (t D_t F - x D_x F - y D_y F)$ \\  \hline
\multicolumn{1}{|l|}{\multirow{1}{*}{$J_3$, Eq.(\protect\ref{CLenergy})}} & \multicolumn{1}{c|}{$\hat{X}_3$} &  $(x H_x + y H_y - t H_t - 5 H) F + H (x D_x F + y D_y F - t D_t F)$ \\  \cline{2-3} 
\multicolumn{1}{|l|}{} & \multicolumn{1}{c|}{$\hat{X}_\infty$} &  $(y f^\prime + f H - g) F + f H D_x F$ \\  \hline
\multicolumn{1}{|l|}{\multirow{1}{*}{$J_4$, Eq.(\protect\ref{CLabsVel})}} & \multicolumn{1}{c|}{$\hat{X}_1$} &  
    $((H_{xxx} + H_{xyy}) H_y - (\beta + H_{xxy} + H_{yyy}) H_x) \Phi^{\prime\prime} F + \Phi^{\prime\prime} F^2 + \Phi^{\prime} D_t F$ \\  \cline{2-3} 
\multicolumn{1}{|l|}{} & \multicolumn{1}{c|}{$\hat{X}_3$} &  
    $\begin{array}{c}
       \left\{
            \left(
                t \left(
                    (H_{xxx}+H_{xyy})H_y
                    -(\beta+H_{xxy}+H_{yyy})H_x
                \right) 
                - y (H_{xxy}+H_{yyy})
                \right. \right.
                \\
                \left. \left.
                - \,x (H_{xyy}+H_{xxx})
                +H_{xx}+H_{yy}
            \right) \Phi^{\prime\prime}
            + 2 \Phi^{\prime}
        \right\} F 
        \\
        + \,t \Phi^{\prime\prime} F^2 
        + \Phi^{\prime} (t D_t F - x D_x F - y D_y F)
     \end{array}$ \\  \cline{2-3} 
\multicolumn{1}{|l|}{} & \multicolumn{1}{c|}{$\hat{X}_\infty$} &  
    $(H_{xxx}+H_{xyy}) f \Phi^{\prime\prime}F + f \Phi^{\prime} D_x F$ 
    \\  \hline
\multicolumn{3}{|c|}{$\beta = 0$}        \\ \hline
\multicolumn{1}{|l|}{\multirow{1}{*}{$J_1$, Eq.(\protect\ref{CLmomentum})}} & \multicolumn{1}{c|}{$\hat{X}^0_3$} &  $\tau_4^\prime(x D_y F - y D_x F)$ \\ \cline{2-3}
\multicolumn{1}{|l|}{} & \multicolumn{1}{c|}{$\hat{X}^0_5$} &  $t \tau_4^\prime(x D_y F - y D_x F)$ \\ \cline{2-3}
\multicolumn{1}{|l|}{} & \multicolumn{1}{c|}{$\hat{X}^0_\infty$} &  $ \tau_4^\prime(\varphi D_x F + \psi D_y F)$ \\ \hline
\multicolumn{1}{|l|}{\multirow{1}{*}{$J_2$, Eq.(\protect\ref{CLnoname1})}} & \multicolumn{1}{c|}{$\hat{X}^0_3$} &  $y \tau_3(x D_y F - y D_x F)$ \\ \cline{2-3}
\multicolumn{1}{|l|}{} & \multicolumn{1}{c|}{$\hat{X}^0_5$} &  $t y \tau_3(x D_y F - y D_x F)$ \\ \cline{2-3}
\multicolumn{1}{|l|}{} & \multicolumn{1}{c|}{$\hat{X}^0_\infty$} &  $ y \tau_3 (\varphi D_x F + \psi D_y F)$ \\ \hline
\multicolumn{1}{|l|}{\multirow{1}{*}{$J_3$, Eq.(\protect\ref{CLenergy})}} & \multicolumn{1}{c|}{$\hat{X}^0_3$} &  $(x H_y - y H_x) F + H (x D_y F - y D_x F)$ \\ \cline{2-3}
\multicolumn{1}{|l|}{} & \multicolumn{1}{c|}{$\hat{X}^0_5$} &  $(2 t (x H_y - y H_x) - x^2 - y^2)F + 2 t H (x D_y F - y D_x F)$ \\ \cline{2-3}
\multicolumn{1}{|l|}{} & \multicolumn{1}{c|}{$\hat{X}^0_\infty$} &  $(\varphi H_x + \psi H_y + y \varphi^\prime - x \psi^\prime - \chi)F + H (\varphi D_x F + \psi D_y F)$ \\ \hline
\multicolumn{1}{|l|}{\multirow{1}{*}{$J_4$, Eq.(\protect\ref{CLabsVel})}} & \multicolumn{1}{c|}{$\hat{X}^0_3$} &  
    $((H_{xxy} + H_{yyy}) x - (H_{xxx} + H_{xyy}) y) \Phi^{\prime\prime} F + \Phi^\prime (x D_y F - y D_x F)$ \\ \cline{2-3}
\multicolumn{1}{|l|}{} & \multicolumn{1}{c|}{$\hat{X}^0_5$} &  
    $\left(2 - \left((H_{xxy} + H_{yyy}) x - (H_{xxx} + H_{xyy}) y\right) t \right)  \Phi^{\prime\prime} F - t \Phi^\prime (x D_y F - y D_x F)$ \\ \cline{2-3}
\multicolumn{1}{|l|}{} & \multicolumn{1}{c|}{$\hat{X}^0_\infty$} &  
    $\left((H_{xxy} + H_{yyy}) \psi + (H_{xxx} + H_{xyy}) \varphi \right) \Phi^{\prime\prime} F + \Phi^\prime (\varphi D_x F + \psi D_y F)$ \\ \hline
\multicolumn{1}{|l|}{\multirow{1}{*}{$J_5^0$, Eq.(\protect\ref{CLb0_1})}} & \multicolumn{1}{c|}{$\hat{X}^0_2$} &  $x\tau_2(2 F + t D_t F)$ \\ \cline{2-3}
\multicolumn{1}{|l|}{} & \multicolumn{1}{c|}{$\hat{X}^0_3$} &  $x \tau_2 (x D_y F - y D_x F)$ \\ \cline{2-3}
\multicolumn{1}{|l|}{} & \multicolumn{1}{c|}{$\hat{X}^0_4$} &  $x \tau_2 (x D_x F + y D_y F)$ \\ \cline{2-3}
\multicolumn{1}{|l|}{} & \multicolumn{1}{c|}{$\hat{X}^0_5$} &  $ t x \tau_2 (x D_y F - y D_x F)$ \\ \cline{2-3}
\multicolumn{1}{|l|}{} & \multicolumn{1}{c|}{$\hat{X}^0_\infty$} &  $ x \tau_2 (\varphi D_x F + \psi D_y F)$ \\ \hline
\multicolumn{1}{|l|}{\multirow{1}{*}{$J_6^0$, Eq.(\protect\ref{CLb0_2})}} & \multicolumn{1}{c|}{$\hat{X}^0_2$} & $(x^2 + y^2) (F + t D_t F)$ \\ \cline{2-3}
\multicolumn{1}{|l|}{} & \multicolumn{1}{c|}{$\hat{X}^0_3$} &  $(x^2 + y^2) (x D_y F - y D_x F)$ \\ \cline{2-3}
\multicolumn{1}{|l|}{} & \multicolumn{1}{c|}{$\hat{X}^0_4$} &  $(x^2 + y^2) (x D_x F + y D_y F)$ \\ \cline{2-3}
\multicolumn{1}{|l|}{} & \multicolumn{1}{c|}{$\hat{X}^0_5$} &  $t (x^2 + y^2) (x D_y F - y D_x F)$ \\ \cline{2-3}
\multicolumn{1}{|l|}{} & \multicolumn{1}{c|}{$\hat{X}^0_\infty$} &  $ (x^2 + y^2) (\varphi D_x F + \psi D_y F)$ \\ \hline
\caption{The derived conservation laws of order three. The conservation laws are given in terms of their multipliers (see (\protect\ref{CLExtForm})).}
\label{tab:cls}\\
\end{longtable}
\egroup

Obviously, the described process can be continued. Applying the generators in characteristic form to the known conservation laws, we will again obtain conservation laws that are, generally, of higher order. The independence of such conservation laws should be studied separately, and we do not dwell on this here.

\section{Conclusions}

The results presented in this publication address some gaps in previous considerations of the geopotential forecast equation.
First, this concerns the group foliations of~(\ref{EqMain0}) (or~(\ref{EqMain}), which is simply another equivalent way of writing equation~(\ref{EqMain0})). For the cases $\beta \neq 0$ and $\beta = 0$, we separately considered the foliations of equation~(\ref{EqMain0}), along with the operators of invariant differentiation, bases of differential invariants, and representations of the equation in terms of these bases and operators. Further consideration of the case~$\beta = 0$ turned out to be unproductive, and we focused on the case $\beta \neq 0$.
For this case, it was shown that the resolving system of equations is consistent and in involution.

At the beginning of Section~\ref{sec:sec3} we discussed some of the advantages of the group foliation approach.
Next, for the case $\beta \neq 0$, we considered invariant solutions corresponding to a fairly simple optimal system of subalgebras of a three-dimensional~Lie algebra. To do this, the group foliation method involves finding solutions to reductions of the resolving system. As it seemed problematic to completely integrate these reductions, we considered some particular solutions, leading to invariant solutions of equation~(\ref{EqMain0}). The intersection of these findings with previously known results obtained by other authors is discussed in detail in Section~\ref{sec:disc}. Many of the solutions we derived generalize previously known results and, apparently, some invariant solutions were obtained for the first time. 
The list of the derived particular solutions is given in Table~\ref{tab:invsols}.
We note that, in the context of group foliations, we haven't exhausted the list of all possible solutions of the resolving system, suggesting the possibility of further research into invariant solutions.

In addition, in Section~\ref{sec:numsol} we considered a case when the resolving system can be solved only numerically, and discussed the difficulties specific to the group foliation approach.

\smallskip

In the final part of the paper, an exhaustive list of second-order conservation laws of equation~(\ref{EqMain0}) was obtained by direct calculations.
Some of these conservation laws are given a physical interpretation. Additionally, based on the second-order conservation laws obtained, a number of third-order conservation laws were also derived using known symmetries. These third-order conservation laws are presented in Table~\ref{tab:cls}.

\smallskip

The listed results contribute to the field of geopotential forecasting, 
complementing previous research related to the symmetry analysis of equation~(\ref{EqMain0}).

\section*{Acknowledgements}

E.I.K. acknowledges Suranaree University of Technology (SUT) and Thailand 
Science Research and Innovation (TSRI) for Full-Time Doctoral Researcher Fellowship.
The author is also grateful to Prof. V.~A.~Dorodnitsyn, Prof. S.~V.~Meleshko, and Prof. O.~V.~Kaptsov for valuable discussions and support over the years.


\begin{thebibliography}{37}
\providecommand{\natexlab}[1]{#1}
\providecommand{\url}[1]{\texttt{#1}}
\expandafter\ifx\csname urlstyle\endcsname\relax
  \providecommand{\doi}[1]{doi: #1}\else
  \providecommand{\doi}{doi: \begingroup \urlstyle{rm}\Url}\fi

\bibitem[Kibel'(1957)]{bk:Kibel1957}
I.~A. Kibel'.
\newblock \emph{Introduction to Hydrodynamic Methods of Short-Term Weather
  Forecasts}.
\newblock Gostekhizdat, Moscow, 1957.
\newblock (in Russian).

\bibitem[Wallace and Hobbs(2006)]{bk:WallaceHobbs[2006]}
J.M. Wallace and P.~V. Hobbs.
\newblock \emph{Atmospheric Science, Second Edition: An Introductory Survey}.
\newblock International Geophysics. Academic Press, 2 edition, 2006.
\newblock ISBN 012732951X; 9780127329512.

\bibitem[Ovsiannikov(1978)]{bk:Ovsiannikov1978}
L.~V. Ovsiannikov.
\newblock \emph{Group Analysis of Differential Equations}.
\newblock Nauka, Moscow, 1978.
\newblock {E}nglish translation, {A}mes, {W}.{F}., Ed., published by Academic
  Press, New York, 1982.

\bibitem[Olver(1986)]{bk:Olver[1986]}
P.~J. Olver.
\newblock \emph{Applications of {L}ie Groups to Differential Equations}.
\newblock Springer-Verlag, New York, 1986.

\bibitem[Huang and Lou(2004)]{HUANG2004428}
Fei Huang and S.~Y. Lou.
\newblock Analytical investigation of {R}ossby waves in atmospheric dynamics.
\newblock \emph{Physics Letters {A}}, 320\penalty0 (5):\penalty0 428--437,
  2004.
\newblock ISSN 0375-9601.
\newblock \doi{10.1016/j.physleta.2003.11.056}.

\bibitem[Bihlo and Popovych(2011)]{bk:BihloPop_Geostr_InvSols[2011]}
A.~Bihlo and R.~O. Popovych.
\newblock {{L}ie symmetry analysis and exact solutions of the quasigeostrophic
  two-layer problem}.
\newblock \emph{Journal of Mathematical Physics}, 52\penalty0 (3):\penalty0
  033103, 03 2011.
\newblock ISSN 0022-2488.
\newblock \doi{10.1063/1.3567175}.

\bibitem[Bihlo and Popovych(2009{\natexlab{a}})]{Bihlo_2009a}
A.~Bihlo and R.~O. Popovych.
\newblock Symmetry analysis of barotropic potential vorticity equation.
\newblock \emph{Communications in Theoretical Physics}, 52\penalty0
  (4):\penalty0 697--700, oct 2009{\natexlab{a}}.
\newblock \doi{10.1088/0253-6102/52/4/27}.

\bibitem[Bihlo and Popovych(2009{\natexlab{b}})]{Bihlo_2009b}
A.~Bihlo and R.~O. Popovych.
\newblock {L}ie symmetries and exact solutions of the barotropic vorticity
  equation.
\newblock \emph{Journal of Mathematical Physics}, 50\penalty0 (12), dec
  2009{\natexlab{b}}.
\newblock \doi{10.1063/1.3269919}.

\bibitem[Fei(2004)]{Fei_2004}
Huang Fei.
\newblock Similarity reductions of barotropic and quasi-geostrophic potential
  vorticity equation.
\newblock \emph{Communications in Theoretical Physics}, 42\penalty0
  (6):\penalty0 903, dec 2004.
\newblock \doi{10.1088/0253-6102/42/6/903}.

\bibitem[Xiao-Yan and Shukla(2008)]{TangXiao-Yan_2008}
Tang Xiao-Yan and Padma~Kant Shukla.
\newblock A note on similarity reductions of barotropic and quasi-geostrophic
  potential vorticity equation.
\newblock \emph{Communications in Theoretical Physics}, 49\penalty0
  (1):\penalty0 229, jan 2008.
\newblock \doi{10.1088/0253-6102/49/1/47}.

\bibitem[Katkov(1965)]{bk:Katkov_Geostr[1965]}
L.~V. Katkov.
\newblock {A class of exact solutions of the equation for the forecast of the
  geopotential}.
\newblock \emph{Izv. Acad. Sci. USSR Atmospher. Ocean. Phys.}, 1:\penalty0
  630–631, 1965.

\bibitem[Katkov(1966)]{bk:Katkov_Geostr[1966]}
L.~V. Katkov.
\newblock {Exact solutions of the geopotential forecast equation}.
\newblock \emph{Akad. Nauk SSSR Ser. Fiz. Atmosfer.~i~Okeana}, 2:\penalty0
  1193, 1966.

\bibitem[Ibragimov(1995)]{bk:HandbookLie_v2}
N.~H. Ibragimov, editor.
\newblock \emph{{CRC} Handbook of {L}ie Group Analysis of Differential
  Equations}, volume~2.
\newblock CRC Press, Boca Raton, 1995.
\newblock N. H. Ibragimov ({\it ed.}).

\bibitem[Ovsyannikov(1970)]{bk:Ovsyannikov_strat[1970]}
L.V. Ovsyannikov.
\newblock Group foliation of the boundary layer equations.
\newblock \emph{Sib. otd. Nauka, Continuum dynamics}, 1, 1970.
\newblock (in Russian).

\bibitem[Vereshchagina(1974)]{bk:Vereshchagina_strat[1974]}
L.I. Vereshchagina.
\newblock Group separation of the spatial nonstationary boundary layer
  equations.
\newblock \emph{Vestnik Leningr. Univ.}, 3\penalty0 (13):\penalty0 82--86,
  1974.
\newblock (in Russian).

\bibitem[Martina et~al.(2001)Martina, Sheftel, and
  Winternitz]{bk:WinternitzMartina_strat[2001]}
L.~Martina, M.~B. Sheftel, and P.~Winternitz.
\newblock Group foliation and non-invariant solutions of the heavenly equation.
\newblock \emph{Journal of Physics {A}: Mathematical and General}, 34\penalty0
  (43):\penalty0 9243, oct 2001.
\newblock \doi{10.1088/0305-4470/34/43/310}.

\bibitem[Nutku and Sheftel'(2001)]{bk:NitkuSheftel_strat[2001]}
Y.~Nutku and M.~B. Sheftel'.
\newblock Differential invariants and group foliation for the complex
  {M}onge--{A}mp{\`e}re equation.
\newblock \emph{Journal of Physics {A}: Mathematical and General}, 34\penalty0
  (1):\penalty0 137, jan 2001.
\newblock \doi{10.1088/0305-4470/34/1/311}.

\bibitem[Golovin(2003)]{bk:Golovin_strat[2003]}
S.~V. Golovin.
\newblock Group stratification and exact solutions of the equation of transonic
  gas motions.
\newblock \emph{Journal of Applied Mechanics and Technical Physics},
  44\penalty0 (3):\penalty0 344--354, May 2003.
\newblock ISSN 1573-8620.
\newblock \doi{10.1023/A:1023429106284}.

\bibitem[Golovin(2004)]{bk:Golovin_strat[2004inv]}
S.~V. Golovin.
\newblock Applications of the differential invariants of infinite dimensional
  groups in hydrodynamics.
\newblock \emph{Communications in nonlinear science and numerical simulations},
  9\penalty0 (1):\penalty0 35--51, 2004.

\bibitem[Chang-Zheng and Shun-Li(2005)]{bk:ChangZheng_start[2005]}
Qu~Chang-Zheng and Zhang Shun-Li.
\newblock Group foliation method and functional separation of variables to
  nonlinear diffusion equations.
\newblock \emph{Chinese Physics Letters}, 22\penalty0 (7):\penalty0 1563, jul
  2005.
\newblock \doi{10.1088/0256-307X/22/7/001}.

\bibitem[Early et~al.(2010)Early, Pohjanpelto, and
  Samelson]{bk:Pohjanpelto_strat[2010]}
J.~J. Early, J.~Pohjanpelto, and R.~M. Samelson.
\newblock Group foliation of equations in geophysical fluid dynamics.
\newblock \emph{Discrete and Continuous Dynamical Systems}, 27\penalty0
  (4):\penalty0 1571--1586, 2010.
\newblock ISSN 1078-0947.
\newblock \doi{10.3934/dcds.2010.27.1571}.

\bibitem[Dorodnitsyn et~al.(2023)Dorodnitsyn, Kaptsov, and
  Meleshko]{bk:DorKapMel_SW2D}
V.~A. Dorodnitsyn, E.~I. Kaptsov, and S.V. Meleshko.
\newblock Lie group symmetry analysis and invariant difference schemes of the
  two-dimensional shallow water equations in {L}agrangian coordinates.
\newblock \emph{Communications in Nonlinear Science and Numerical Simulation},
  119:\penalty0 107119, 2023.
\newblock ISSN 1007-5704.
\newblock \doi{10.1016/j.cnsns.2023.107119}.

\bibitem[Pommaret(1978)]{bk:Pommaret}
J.~E. Pommaret.
\newblock \emph{Systems of partial differential equations and {L}ie
  Pseudogroups}.
\newblock Gordon \& Breach, New York, 1978.

\bibitem[Meleshko(2005)]{bk:Meleshko[2005]}
S.~V. Meleshko.
\newblock \emph{Methods for Constructing Exact Solutions of Partial
  Differential Equations}.
\newblock Mathematical and Analytical Techniques with Applications to
  Engineering. Springer, New York, 2005.

\bibitem[Xu(2008)]{bk:xu2008algebraic}
Xiaoping Xu.
\newblock Algebraic approaches to the geopotential forecast and nonlinear {MHD}
  equations, 2008.

\bibitem[Patera and Winternitz(1977)]{bk:PateraWinternitz1977}
J.~Patera and P.~Winternitz.
\newblock Subalgebras of real three- and four-dimensional {L}ie algebras.
\newblock \emph{Journal of Mathematical Physics}, 18\penalty0 (7):\penalty0
  1449--1455, 1977.
\newblock \doi{10.1063/1.523441}.

\bibitem[Jamal(2020)]{bk:Jamal2020}
S.~Jamal.
\newblock New multipliers of the barotropic vorticity equations.
\newblock \emph{Analysis and Mathematical Physics}, 10\penalty0 (2):\penalty0
  21, Apr 2020.
\newblock ISSN 1664-235X.
\newblock \doi{10.1007/s13324-020-00365-4}.

\bibitem[{Andrews}(1983)]{DGAndrews1983}
D.~G. {Andrews}.
\newblock A conservation law for small-amplitude quasi-geostrophic disturbances
  on a zonally asymmetric basic flow.
\newblock \emph{Journal of the Atmospheric Sciences}, 40\penalty0 (1):\penalty0
  85--90, jan 1983.
\newblock \doi{10.1175/1520-0469(1983)040<0085:ACLFSA>2.0.CO;2}.

\bibitem[Bihlo et~al.(2019)Bihlo, Cardoso-Bihlo, and
  Popovych]{bk:BihloPop_Geostr_numeric[2019]}
A.~Bihlo, E.~Cardoso-Bihlo, and R.~O. Popovych.
\newblock Invariant parameterization of geostrophic eddies in the ocean, 2019.

\bibitem[Kaptsov and Meleshko(2019)]{bk:KaptsovMeleshko2019}
E.~I. Kaptsov and S.~V. Meleshko.
\newblock Conservation laws of the two-dimensional gas dynamics equations.
\newblock \emph{International Journal of Non-Linear Mechanics}, 112:\penalty0
  126--132, 2019.

\bibitem[Kaptsov et~al.(2022)Kaptsov, Meleshko, and
  Dorodnitsyn]{bk:GilmanModelKapMelDor[2022]}
E.~I. Kaptsov, S.~V. Meleshko, and V.~A. Dorodnitsyn.
\newblock Symmetries and conservation laws of the one-dimensional shallow water
  magnetohydrodynamics equations in {L}agrangian coordinates.
\newblock \emph{Journal of Physics A: Mathematical and Theoretical},
  55\penalty0 (49):\penalty0 495202, dec 2022.
\newblock \doi{10.1088/1751-8121/aca84a}.

\bibitem[Dorodnitsyn et~al.(2015)Dorodnitsyn, Kaptsov, Kozlov, and
  Winternitz]{DorKapKozWin2015}
V.~A. Dorodnitsyn, E.~I. Kaptsov, R.~Kozlov, and P.~Winternitz.
\newblock The adjoint equation method for constructing first integrals of
  difference equations.
\newblock \emph{Journal of Physics A: Mathematical and Theoretical},
  48\penalty0 (5):\penalty0 055202, jan 2015.
\newblock \doi{10.1088/1751-8113/48/5/055202}.

\bibitem[Ibragimov(2011)]{bk:Ibragimov2011}
N.~H. Ibragimov.
\newblock Nonlinear self-adjointness and conservation laws.
\newblock \emph{Journal of Physics {A}: Mathematical and Theoretical},
  44\penalty0 (43):\penalty0 432002, oct 2011.
\newblock \doi{10.1088/1751-8113/44/43/432002}.

\bibitem[Anco(2017)]{bk:Anco_AntiIbragimov[2017]}
S.~C. Anco.
\newblock On the incompleteness of {I}bragimov's conservation law theorem and
  its equivalence to a standard formula using symmetries and
  adjoint-symmetries.
\newblock \emph{Symmetry}, 9\penalty0 (3), 2017.
\newblock ISSN 2073-8994.
\newblock \doi{10.3390/sym9030033}.

\bibitem[Anco and Bluman(1997)]{bk:AncoBluman1997}
S.~C. Anco and G.~Bluman.
\newblock Direct construction of conservation laws from field equations.
\newblock \emph{Phys. Rev. Lett.}, 78\penalty0 (3):\penalty0 2869--2873, 1997.
\newblock \doi{10.1103/PhysRevLett.78.2869}.

\bibitem[Bluman et~al.(2010)Bluman, Cheviakov, and
  Anco]{bk:BlumanCheviakovAnco}
G.~W. Bluman, A.~F. Cheviakov, and S.~C. Anco.
\newblock \emph{Applications of Symmetry Methods to Partial Differential
  Equations}.
\newblock Springer, New York, 2010.
\newblock Applied Mathematical Sciences, Vol.168.

\bibitem[Naz et~al.(2008)Naz, Mahomed, and Mason]{bk:NazMahomedMason2008}
R.~Naz, F.~M. Mahomed, and D.~P. Mason.
\newblock Comparison of different approaches to conservation laws for some
  partial differential equations in fluid mechanics.
\newblock \emph{Applied Mathematics and Computation}, 205\penalty0
  (1):\penalty0 212--230, 2008.
\newblock ISSN 0096-3003.
\newblock \doi{10.1016/j.amc.2008.06.042}.
\newblock Special Issue on Life System Modeling and Bio-Inspired Computing for
  {LSMS} 2007.

\end{thebibliography}

\end{document}